\documentclass[fleqn,usenatbib]{mnras}

\usepackage{newtxtext,newtxmath}

\usepackage[T1]{fontenc}
\usepackage{ae,aecompl}

\DeclareRobustCommand{\VAN}[3]{#2}
\let\VANthebibliography\thebibliography
\def\thebibliography{\DeclareRobustCommand{\VAN}[3]{##3}\VANthebibliography}

\usepackage{graphicx}	
\usepackage{amsmath}	
\usepackage{siunitx}
\usepackage{subcaption}
\usepackage{threeparttable}
\usepackage{float}
\usepackage{anyfontsize}
\usepackage{lscape} 

\title{Determination of dynamical ages of open clusters through the A$^+$ parameter -- I}

\author[Khushboo K. Rao et al.]{
Khushboo K. Rao,$^{1}$ \thanks{E-mail: p20170419@pilani.bits-pilani.ac.in} 
Kaushar Vaidya,$^{1}$ 
Manan Agarwal,$^{1}$ 
and Souradeep Bhattacharya$^{2}$
\\
$^{1}$ Department of physics, Birla Institute of Technology and Science-Pilani, 333031 Rajasthan, India\\
$^{2}$ Inter University Centre for Astronomy and Astrophysics, Ganeshkhind, Post Bag 4, Pune 411007, India
}

\date{Accepted 2021 October 4. Received 2021 October 2; in original form 2021 February 9}

\pubyear{2021}

\begin{document}
\label{firstpage}
\pagerange{\pageref{firstpage}--\pageref{lastpage}}
\maketitle

\begin{abstract}
The sedimentation level of blue straggler stars (BSS) has been shown to be a great tool to investigate the dynamical states of globular clusters (GCs). The area enclosed between the cumulative radial distributions of BSS and a reference population up to the half-mass radius of the clusters, $A^+_{\mathrm{rh}}$, is known to be a measure of the sedimentation of BSS in GCs. In this work, we calculate $A^+_{\mathrm{rh}}$ for 11 open clusters (OCs) using a combined list of main-sequence turn-off stars, sub-giant branch stars, and red-giant branch stars as reference population. The BSS, the reference populations, and the cluster members are identified using the proper motions and parallaxes from the Gaia DR2 data. In a subset of clusters, the BSS are confirmed cluster members on the basis of radial velocity information available in the literature. Using the Pearson and Spearman rank correlation coefficients, we find weak correlations between the estimated values of $A^+_{\mathrm{rh}}$ and other markers of dynamical ages of the clusters, i.e., the number of central relaxations a cluster has experienced since its formation, and the structural parameters of the clusters. Based on statistical tests, we find that these correlations are similar to the corresponding correlations among the less evolved GCs, albeit within large errors.
\end{abstract}


\section{Introduction}
A star cluster is a gravitationally bound system of stars having a wide mass spectrum, in which various kinds of stellar interactions occur and lead to several dynamical processes in the cluster environment, such as two-body relaxation, mass segregation due to equipartition of energy, stellar encounters, and binary system evolution, etc. \citep{Meylan1997}. These dynamical processes result in exotic populations like blue straggler stars \citep[BSS;][]{Stryker1993,Bailyn1995}, millisecond pulsars \citep{Bhattacharya1991}, and cataclysmic variables \citep{Ritter2010}.  Among these stellar populations, BSS are particularly interesting because they form in the majority and are commonly found in diverse environments such as globular clusters \citep[hereafter GCs;][]{Sandage1953, Simunovic2016}, open clusters \citep[hereafter OCs;][]{Johnson1955, deMarchi2006}, Galactic fields \citep{Santucci2015}, and dwarf galaxies \citep{Monelli2012}, and are easy to be identified based on their location on color-magnitude diagrams (CMDs). The standard concept of BSS formation is that their progenitor gains additional mass and rejuvenates hydrogen burning in the core, though the theories differ in how the extra mass is acquired. Currently, there are three fundamental theories through which BSS can originate, i) stellar collisions - direct collision between single stars \citep{Hurley2005,Chatterjee2013,Hypki2013}, a stellar collision in dynamical interaction of binaries with single stars or with another binary \citep{Leonard1989, Leigh2019}, ii) the merger of an inner binary in a triple system through the Kozai mechanism \citep{Perets2009, Naoz2014}, iii) mass transfer in a binary system \citep{McCrea1964,Gosnell2014}.

As a star cluster evolves, dynamical friction (hereafter DF) segregates sources in the cluster core based on the descending order of their masses \citep{Chandrasekhar1943}. DF impacts massive stars first and brings them from a more considerable distance to the cluster center to place them in the cluster core, at the same time pushing the lighter stars further away from the cluster center. Since BSS are among the massive members of the cluster populations \citep{Shara1997}, DF starts to influence those BSS first which are nearer to the cluster center, gradually affecting BSS from the peripheral regions, as the cluster evolves. One cannot directly measure the extent of the cluster up to which the DF is effective. \citet{Ferraro2012} plotted double normalized radial distributions of BSS against a reference population (horizontal branch stars in their case) and classified GCs in three distinct families. Family I GCs have flat BSS radial distribution in which DF has not been effective yet to segregate the cluster BSS in its core. These are dynamically young clusters. Family II GCs show bimodal radial distributions, a central peak followed by a minima at a certain radial distance from the cluster center, and a rising trend in the outskirts of the cluster. The minima in the bimodal radial distribution, $r_{\mathrm{min}}$, is a cluster radius up to which the DF is effective such that the BSS up to this radius are mass segregated. The Family II clusters are of intermediate dynamical age. Family III GCs show unimodal BSS radial distributions with a central peak that monotonically decreases throughout the cluster extension.  These are dynamically old clusters in which DF has segregated all the BSS in the cluster core. \citet{Ferraro2012} found a strong anticorrelation between  $r_{\mathrm{min}}$ and the central relaxation time of GCs normalized to the Hubble time, $t_{\mathrm{rc}}/t_{\mathrm{H}}$. Several authors, e.g., \citet{Beccari2013}, \citet{Dalessandro2013}, \citet{Sanna2014}, \citet{Dalessandro2015}, estimated $r_{\mathrm{min}}$ for individual clusters and compared them with other markers of the dynamical age and proved that $r_{\mathrm{min}}$ is a powerful indicator of the dynamical evolution for GCs.

For the first time in the literature, such a study in OCs was undertaken by \citet{Vaidya2020}. They found that even among the OCs, a similar correlation between the two cluster parameters, $N_{\mathrm{relax}}$, defined as the ratio of the cluster age to the central relaxation time ($C_{\mathrm{Age}}/t_{\mathrm{rc}}$), and $r_{\mathrm{min}}$, as previously reported in a large number of GCs, is observed \citep{Ferraro2012}. For such an analysis, however, one needs to be careful in estimating the $r_{\mathrm{min}}$. If the bin size is too large, the error in $r_{\mathrm{min}}$ increases up to the width of the chosen bin size, however, if it is too small, one gets noisy points near $r_{\mathrm{min}}$ because of diminishing numbers of BSS near this region \citep[zone of avoidance;][]{Miocchi2015}.  This fine-tuning of bin size for an accurate determination of $r_{\mathrm{min}}$ becomes particularly challenging for the radial distributions of OCs since they typically contain much fewer BSS compared to GCs.

\citet{Alessandrini2016} performed N-body simulations of GCs with different fractions of dark remnants (neutron stars and black holes), and proposed a new parameter, $A^+$, to measure the sedimentation level of the BSS that indicates the dynamical state of the cluster, where $A^+$ is the area confined between the cumulative radial distributions of the BSS and a reference population. Depending on the available photometric data and clusters properties, different cluster populations such as horizontal branch stars (HBs), red-giant branch stars (RGBs), sub-giant branch stars (SGBs), and main-sequence turn-off stars (MS-TO stars), i.e, the less massive cluster members than the BSS, have been used as a reference population in the literature to estimate the value of $A^+$  \citep{Lanzoni2016,Ferraro2018,Ferraro2020}. Unlike $r_{\mathrm{min}}$, $A^+$ does not require binning of the data \citep{Lanzoni2016,Raso2017,Ferraro2018,Ferraro2020}. As a cluster evolves, BSS start to segregate in the cluster center more rapidly than any reference population, leading to increasing separation between the two cumulative radial distributions. Black holes delay the mass segregation process, though do not prevent BSS from segregation, which implies that $ A^+ $ always increases with time, whether it is a slow or a rapid increment \citep{Alessandrini2016}. 

\citet{Lanzoni2016} performed the first observational estimation of $A^+$ in 25 GCs. This work was extended to $\sim$33$\%$ population of GCs by \citet{Raso2017} and \citet{Ferraro2018},  and to 5 LMC GCs by \citet{Ferraro2020}. OCs are still unexplored systems in this domain, therefore, we pursue a similar study on OCs. OCs and GCs are vastly different stellar systems in terms of their shapes, ages, stellar densities and locations \citep{Janes1982,Lada2010,Harris1979,Freeman1981}. There are certain advantages to study the BSS of OCs.  Because of lower stellar densities, the individual BSS can be studied in detail.  With the Gaia DR2 \citep{Gaia2018} and now Gaia EDR3 \citep{Gaia2020}, the precise information of positions and proper motions for billions of stars, secure cluster members and BSS populations can be identified.  In two OCs, NGC 188 and M67, the BSS population and their binary origin, has been examined with great details that has shed light on the formation channels of BSS \citep{Gosnell2015,Subramaniam2016,Sindhu2019,Jadhav2019}.  Also, \citet{Geller2008} have used N-body simulations to reproduce the observed BSS (binarity, radial distributions), which is only possible in OCs.  More recently, the BSS populations of OCs have been studied to explore the link between the clusters' dynamical status and the observed BSS radial distributions \citep{Bhattacharya2019,Vaidya2020,Rain2020a,Rain2020b}.

In this work, we present our measurements of $A^+$ in 11 OCs whose BSS populations have been identified in the literature. These OCs contain 10 or more BSS. We use a combined list of MS-TO stars, SGBs, RGBs, and red clump stars (RCs) as reference population in this study. The rest of the paper is arranged as follows. In \S \ref{section:Data}, we give information about the sources from which we have taken the list of member stars of the 11 OCs and their BSS candidates or confirmed BSS. In \S \ref{section:selection-criteria}, we establish selection criteria for BSS, MS-TO stars, SGBs, RGBs, and RCs. In \S \ref{section:Calculation-of-A}, we give details of the calculation of $A^+$ and its error estimation for the 11 OCs. In \S \ref{section:Trc}, we discuss the procedure to estimate the central relaxation time. In \S \ref{section:Results}, we discuss the relations of $A^+$ with other markers of clusters' dynamical age, and compare those with GCs. In the end, \S \ref{section:Summary} concludes the present work.

\section{Target Clusters and BSS Samples} 
\label{section:Data}
\begin{table}
	\centering
	\caption{The list of clusters studied in this work, their ages, distances, the number of BSS, the number of REF, and $G_{\mathrm{TO}}$.}
	\label{tab:Table1}
	\begin{tabular}{cccccc} 
		\hline
		\\
		Cluster & Age & Dist & No. of & No. of & $G_{\mathrm{TO}}$  \\
		 & (Gyr) & (pc) & BSS & REF &  \\
		\\
		\hline
		\\
		Berkeley 17$^{a}$ & 10 & 3138.6 & 14 & 47 & 18.05 \\
		Berkeley 39$^{b}$ & 6 & 4254 & 16 & 178 & 17.00  \\
		Collinder 261$^{e*}$ & 6 & 3053 & 43 & 431 & 16.67 \\
		NGC 188$^{b}$ & 7 & 1700 & 15 & 195 & 15.181 \\
		NGC 2158$^{b}$ & 2.2 & 4254 & 36 & 402 & 16.78 \\
		NGC 2506$^{b}$ & 2.2 & 3110 & 10 & 268 & 15.319 \\
		NGC 2682$^{c*}$ & 4  & 850 & 13 & 191 & 13.291 \\
		NGC 6791$^{b}$ & 8.4  & 4475 & 25 & 274 & 17.507 \\
		NGC 6819$^{b}$ & 2.73  & 2652 & 14 & 269 & 15.34 \\
		NGC 7789$^{d*}$ & 1.9 & 1965 & 12 & 546 & 14.583 \\
		Trumpler 5$^{f*}$ & 3.4 & 3226 & 35 & 324 & 16.728$^{\#}$ \\
		\\
		\hline
	\end{tabular}
	\begin{tablenotes}
		\item {BSS used to calculate $A^{+}_{\mathrm{rh}}$ are taken from:}
		\item {$^a$\citet{Bhattacharya2019}; $^b$\citet{Vaidya2020}; $^c$\citet{Geller2015}; $^d$\citet{Nine2020}; $^e$\citet{Rain2020a}; $^f$\citet{Rain2020b}}
		\item {ages and distances are taken from:}
		\item {$^a$\citet{Bhattacharya2019}; $^b$\citet{Vaidya2020}; $^*$the references are listed in Table \ref{tab:Table_A2}}
		\item $^{\#}$obtained after the differential reddening correction
	\end{tablenotes}
\end{table}
Gaia is a space mission to map the whole sky in three dimensions. The Gaia DR2 data \citep{Gaia2018} provides five parameters astrometric solution - positions (${\alpha}$, ${\delta}$), parallaxes, proper motions for more than 1.3 billion stars of the Galaxy and throughout the local group of galaxies. This dataset has unparalleled precision in parallaxes and proper motions. In parallaxes, the uncertainties are of the order 0.4 milliarcsecond (hereafter mas) for G < 15 mag, 0.1 mas for G = 17 mag, and 0.7 mas at the faint end, G = 20 mag. In proper motions, the corresponding uncertainties are up to 0.06 mas yr$^{-1}$ for G < 15 mag, 0.2 mas yr$^{-1}$ for G = 17 mag, and 1.2 mas yr$^{-1}$ for G = 20 mag \citep{Gaia2018}.
\begin{figure*}
    \centering
	\begin{subfigure}[b]{0.32\textwidth}
    		\includegraphics[width=1.0\textwidth]{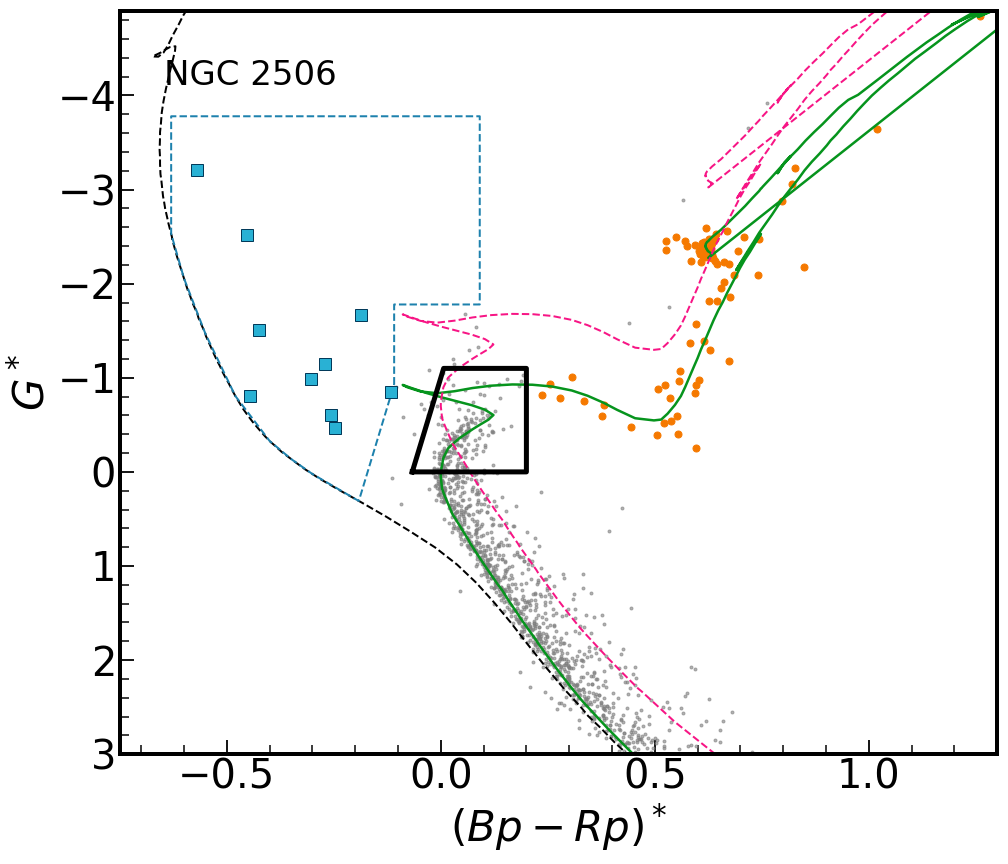}
		\caption*{}
	\end{subfigure}
	\quad 
	\begin{subfigure}[b]{0.32\textwidth}
   		\includegraphics[width=1.0\textwidth]{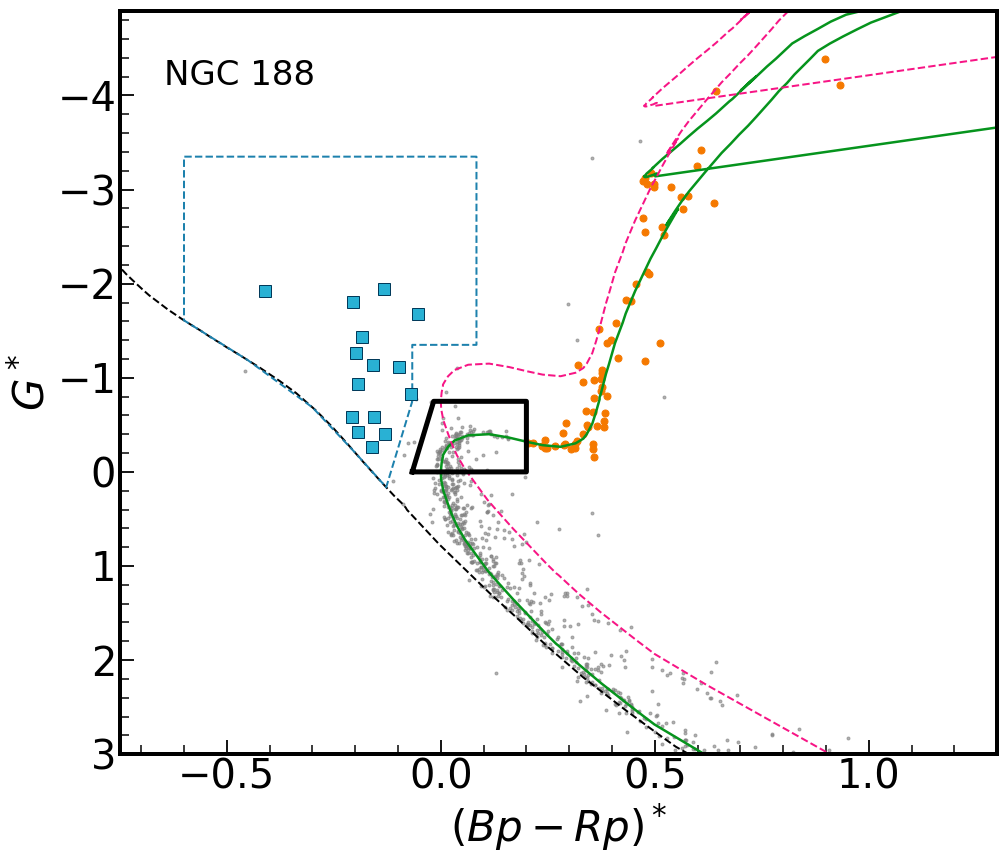}
		\caption*{}
	\end{subfigure}
	\quad 
	\begin{subfigure}[b]{0.32\textwidth}
		\includegraphics[width=1.0\textwidth]{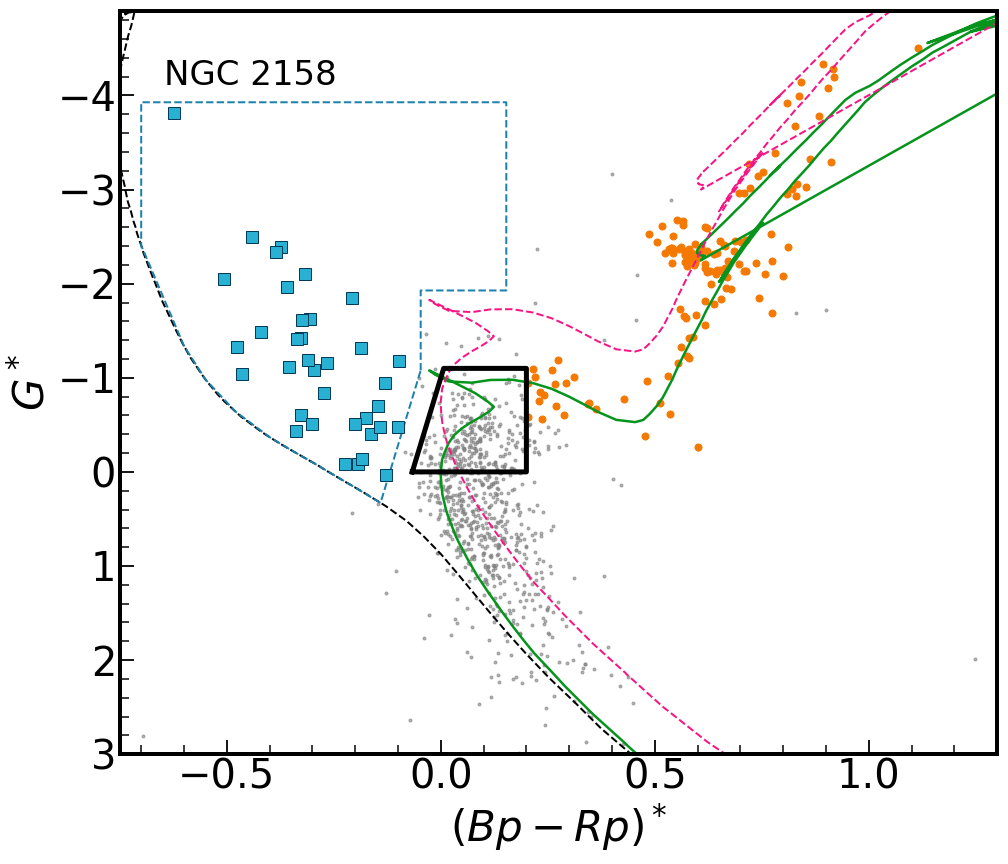}
		\caption*{}
	\end{subfigure}
	\quad 
	\begin{subfigure}[b]{0.32\textwidth}
   		\includegraphics[width=1.0\textwidth]{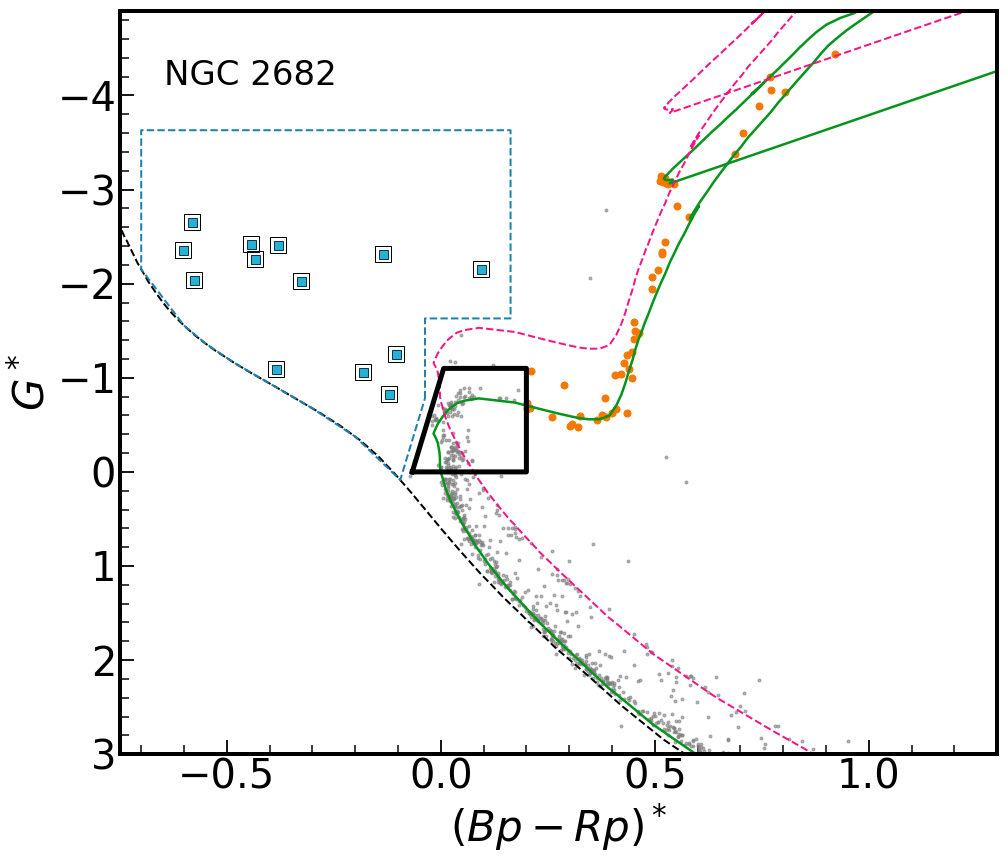}
		\caption*{}
	\end{subfigure}
	\quad
	\begin{subfigure}[b]{0.32\textwidth}
    		\includegraphics[width=1.0\textwidth]{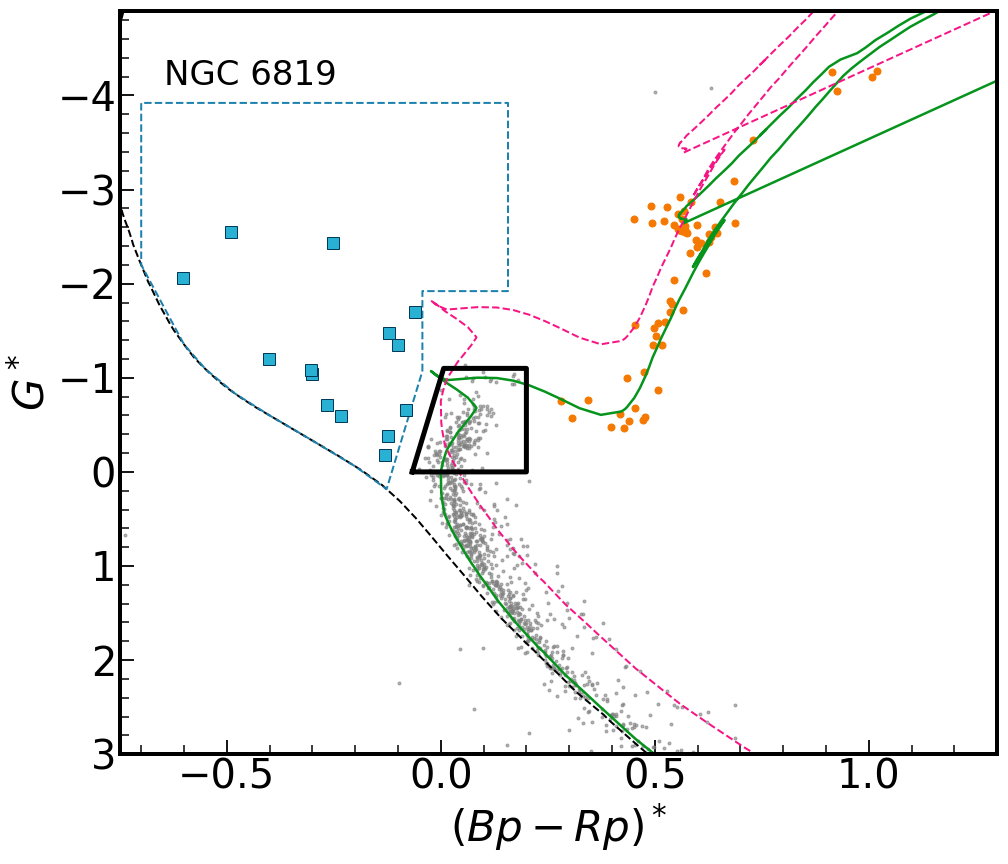}
		\caption*{}
	\end{subfigure}
	\quad 
	\begin{subfigure}[b]{0.32\textwidth}
   		\includegraphics[width=1.0\textwidth]{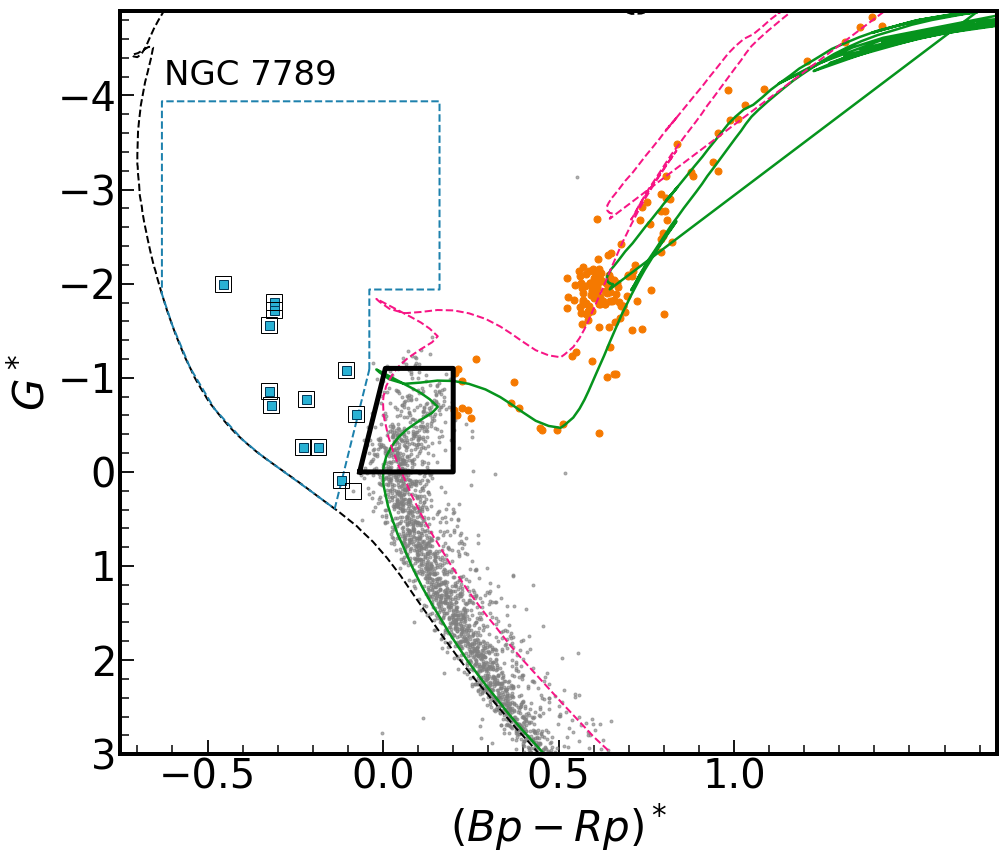}
		\caption*{}
	\end{subfigure}
	\quad
	\begin{subfigure}[b]{0.32\textwidth}
		\includegraphics[width=1.0\textwidth]{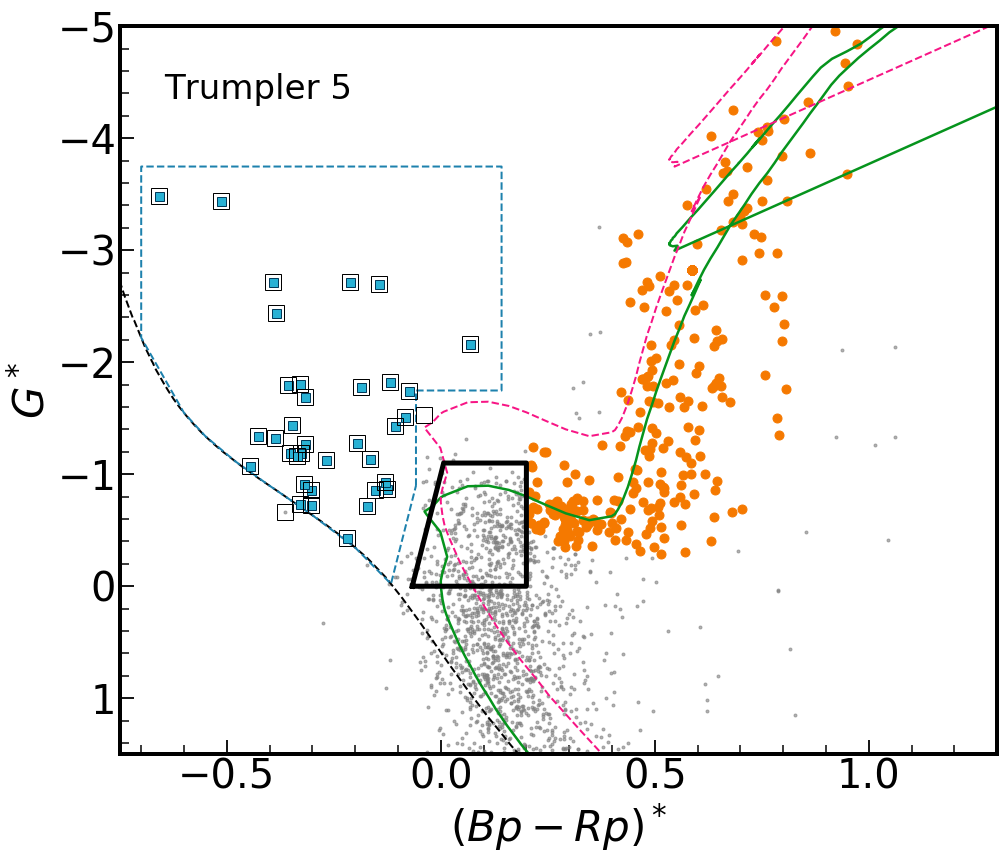}
		\caption*{}
	\end{subfigure}
	\quad
	\begin{subfigure}[b]{0.32\textwidth}
    		\includegraphics[width=1.0\textwidth]{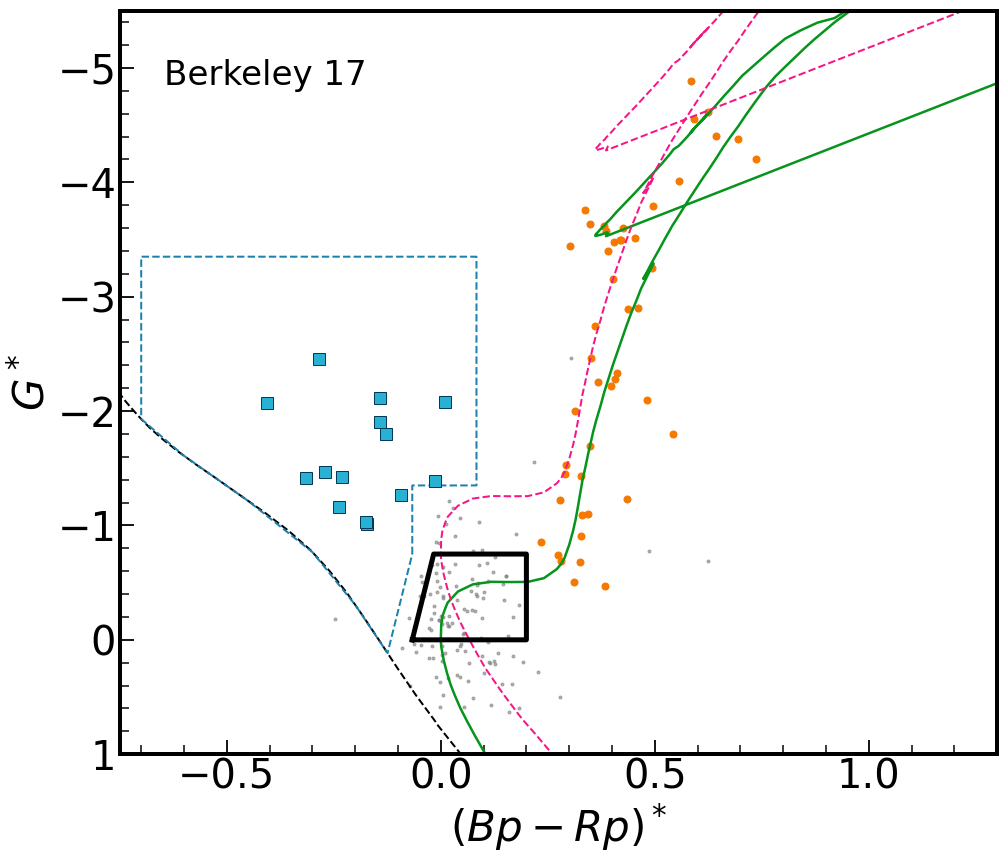}
		\caption*{}
	\end{subfigure}
	\quad 
	\begin{subfigure}[b]{0.32\textwidth}
   		\includegraphics[width=1.0\textwidth]{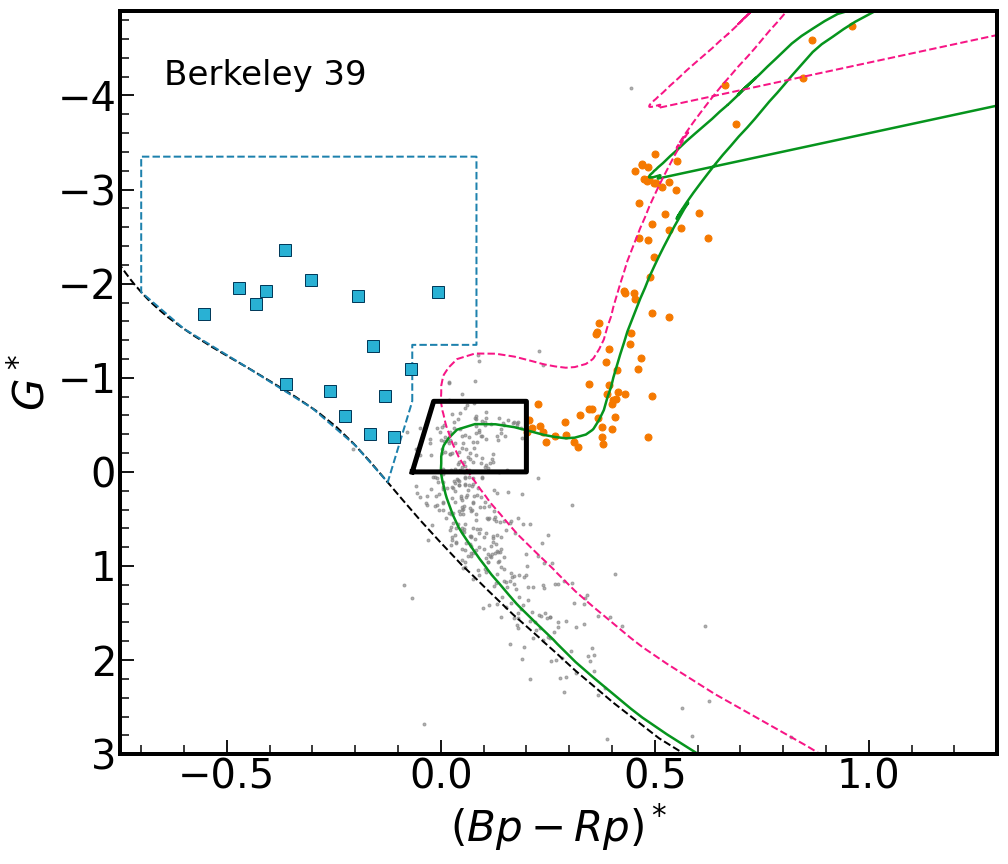}
		\caption*{}
	\end{subfigure}
	\quad
	\begin{subfigure}[b]{0.32\textwidth}
		\includegraphics[width=1.0\textwidth]{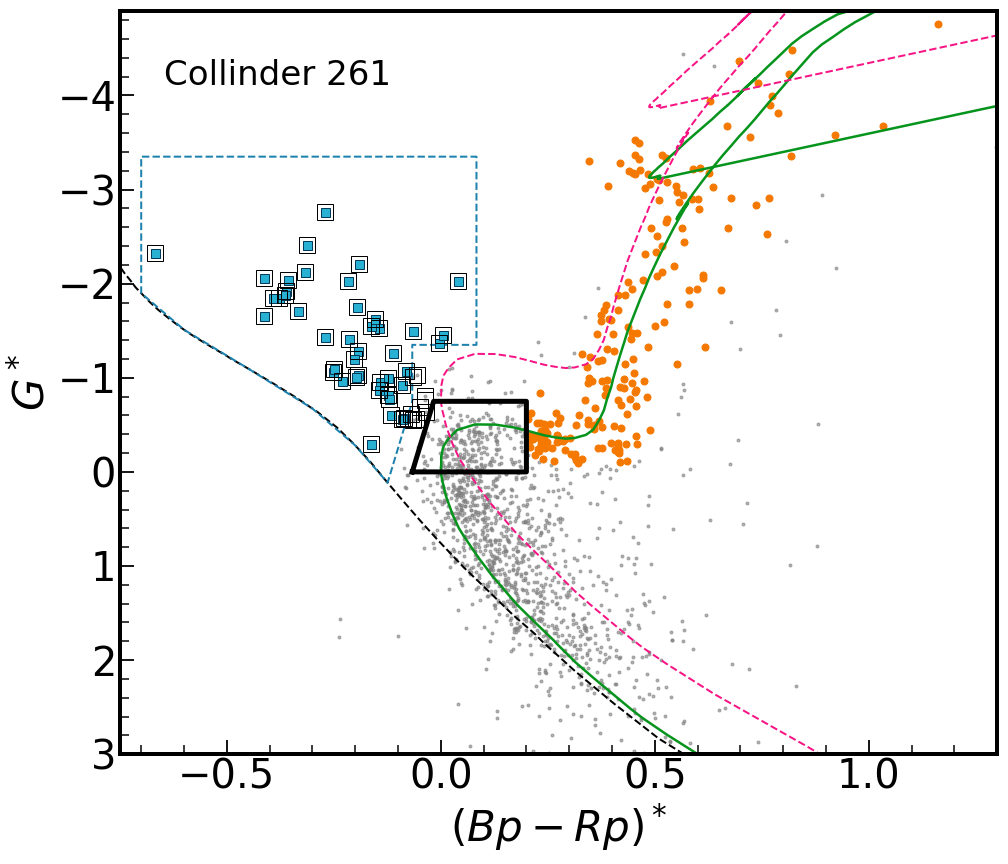}
		\caption*{}
	\end{subfigure}
	\quad
	\begin{subfigure}[b]{0.32\textwidth}
		\includegraphics[width=1.0\textwidth]{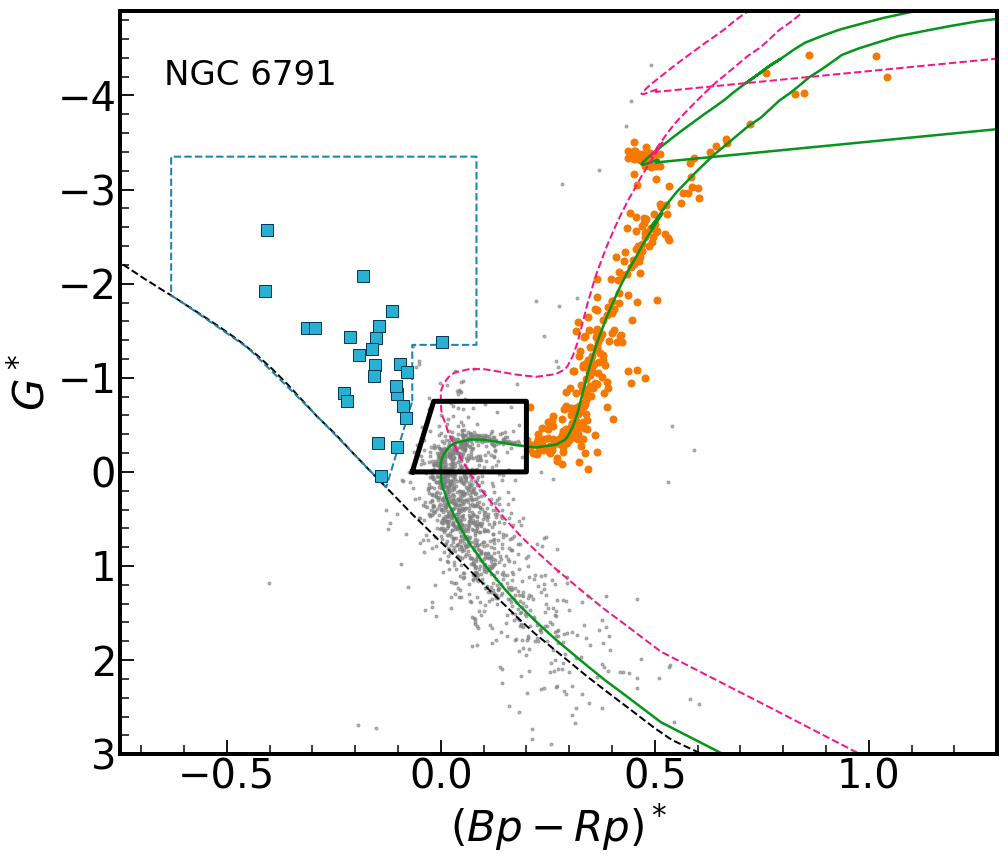}
		\caption*{}
	\end{subfigure}
	\caption{Normalized CMDs of 11 OCs with fitted PARSEC isochrones (green curve), PARSEC isochrones shifted up by 0.75 mag (magenta dashed curve), and zero-age main-sequence (black dashed curve). BSS candidates identified using our BSS selection criteria (blue dashed box) are shown as blue filled squares. BSS taken from the literature are shown as black open squares. The black trapezoid represents the MS-TO stars selection box. SGBs, RGBs, and RCs are shown as orange filled circles. Main-sequence stars and other remaining cluster members are shown as grey dots.}
	\label{fig:Figure 1}
\end{figure*}
\begin{figure*}
	\includegraphics[width=1.0\textwidth]{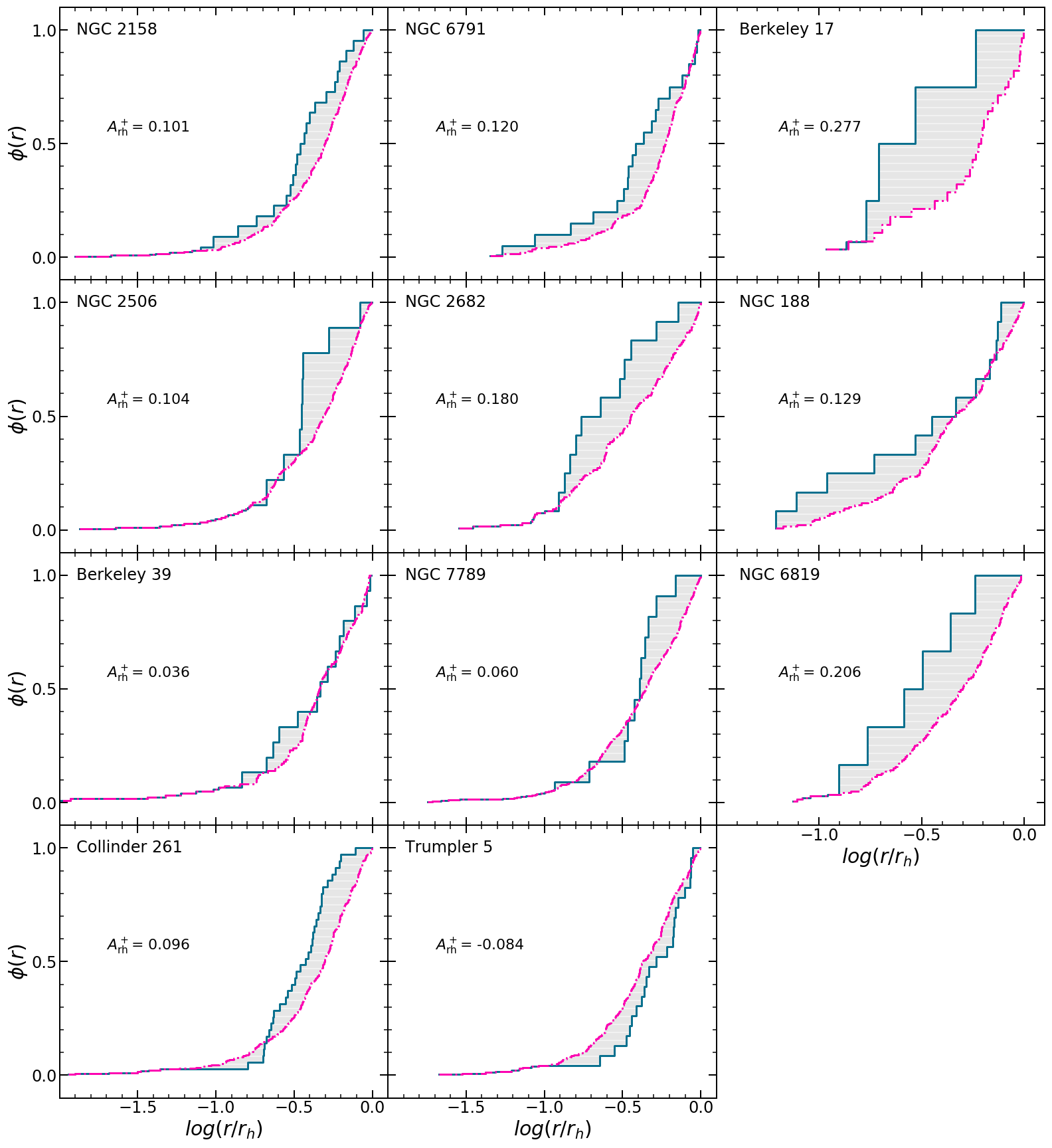}
    \caption{The cumulative radial distributions of the BSS (blue curve) and the REF population (magenta dashed curve), plotted against logarithm of the radial distance from the cluster center in the units of $r_{\mathrm{h}}$,  of the 11 OCs. For Berkeley 17, NGC 6791, and Trumpler 5, SGBs, RGBs, and RCs, whereas, for remaining clusters, REF is MS-TO stars, SGBs RGBs, and RCs. See section \ref{section:selection-criteria} for details. The values of $A^+_{\mathrm{rh}}$ shown on each plot correspond to the grey shaded portion between the cumulative radial distributions of the BSS and the REF population.}
    	\label{fig:Figure 2}
\end{figure*}
The present study focuses on studying the correlation of $A^+$ with the clusters' structural parameters and with the theoretical estimate of the relaxation status of the clusters. For this purpose, it is imperative that we choose OCs which contain a reasonably large number of BSS that allow such an analysis. Since most OCs contain small numbers of BSS, finding such OCs is not an easy task. \citet{Vaidya2020} studied 7 OCs, Berkeley 39, Melotte 66, NGC 188, NGC 2158, NGC 2506, NGC 6791, and NGC 6819 which contain a minimum of 14 BSS, using the Gaia DR2 data to identify cluster members and BSS candidates. Some of these clusters were previously part of WIYN Open Cluster Survey (WOCS) and had confirmed BSS, i.e., NGC 188 \citep{Geller2008}, NGC 6791 \citep{Tofflemire2014}, and  NGC 6819 \citep{Milliman2014}. We choose 6 clusters, Berkeley 39, NGC188, NGC2158, NGC2506, NGC6791, and NGC 6819, out of these seven clusters in our present work. In addition, there are 5 clusters that we include in this work. For three of them, the BSS populations have been identified using the Gaia DR2 in recent studies. \citet{Bhattacharya2019} studied BSS populations of Berkeley 17. \citet{Rain2020a} studied the BSS of Collinder 261, and \citet{Rain2020b} studied the BSS of Trumpler 5. In both, Collinder 261 and Trumpler 5, \citet{Rain2020a,Rain2020b} also had a small number of BSS with spectra from the high-resolution spectrograph FLAMES/GIRAFFE@VLT. In two additional OCs, NGC 2682 and NGC 7789, the complete BSS populations as well as a large number of MS-TO stars have been studied spectroscopically under the WIYN Open Cluster Survey \citep{Geller2015,Nine2020}. We adopt the BSS samples of these 11 clusters from these studies, and apply our BSS selection criteria on these samples, as described in detail in $\S$ \ref{section:selection-criteria}. For the seven clusters from \citet{Vaidya2020}, we already have the cluster members. For Berkeley 17, we get the cluster members from \citet{Bhattacharya2019}.\\
\indent For the remaining four clusters, i.e., Collinder 261, Trumpler 5, NGC 2682, and NGC 7789, we identify the cluster members using the Gaia DR2 in order to select the MS-TO stars, SGBs, and RGBs as the reference population, whereas the BSS lists are taken from the literature. The membership identification procedure and the related figures for the four clusters are presented in Appendix \ref{Appendix}. Figure \ref{fig:Figure A5} shows the observed CMDs of the 11 OCs, where the CMD shown for Trumpler 5 is obtained after the differential reddening correction. Among these four clusters, Trumpler 5 has a broader main-sequence and elongated red-clump stars, i.e., the cluster is highly affected by the differential presence of the dust along its line of sight. Therefore we perform differential reddening correction in the identified members of the cluster (as also carried out by \citet{Rain2020b}). The differential reddening correction procedure is explained in Appendix \ref{section:Dr-Tr5} and the related figures are shown in Figure \ref{fig:Figure B1}. The differential reddening effect is negligible in the other clusters. Table \ref{tab:Table1} lists the 11 OCs that we study in this work with their fundamental properties such as age and distance.

\section{The BSS and reference population Selection}
\label{section:selection-criteria}
As we want to compare the BSS with the reference population, the reference population needs a similar completeness as the BSS. The Gaia DR2 data is essentially complete between G = 12 to G = 17 mag \citep{Gaia2018}. The completeness declines for faint stars relative to bright stars in denser regions. The crowding limit for the G photometry and astrometry is 1050000 sources/deg$^{2}$, whereas for the BP/RP photometry it is 750000 sources/deg$^{2}$ \citep{Gaia2016}. The BP/RP photometry has lower source density, therefore, the effects of crowding are even more essential to incorporate in this subset. Hence, to ensure that our analysis is not affected by the incompleteness in Gaia DR2 data, we have selected reference populations according to $G_{\mathrm{TO}}$  of the clusters, where $G_{\mathrm{TO}}$ is the G mag of MS-TO point. The bright stars with G < 16 mag are not affected by incompleteness. Therefore, for the clusters NGC 188, NGC 2506, NGC 2682, NGC 6819, and NGC 7789, whose $G_{\mathrm{TO}}$ < 16 mag, we combine MS-TO stars with SGBs, RGBs, and \textit{RCs} as reference populations. In Berkeley 39,  Collinder 261, and NGC 2158, the $G_{\mathrm{TO}}$ ranges between G = 16 and G = 17 mag. Hence, in order to gauge the seriousness of the effect of crowding in these clusters, we estimate the number density of sources around the centers of these clusters. We find these values as: Berkeley 39 -- $2.92 \times 10^4$ sources/deg$^2$, Collinder 261 -- $4.26 \times 10^5$ sources/deg$^2$, and NGC 2158 -- $4.92 \times 10^4$ sources/deg$^2$. Since all the three clusters have number densities less than the crowding limit of BP/RP photometry, we include MS-TO stars in the list of SGBs, RGBs, and RCs to use them as a reference population for these clusters. For Berkeley 17, and NGC 6791 which have $G_{\mathrm{TO}}$ > 17 mag, we use only SGBs, RGBs, and RCs as a reference population. Since the observed $G_{\mathrm{TO}}$ of Trumpler 5 is greater than G = 17 mag, we use SGBs, RGBs, and RCs as the reference population. The $G_{\mathrm{TO}}$ and the number of reference population of each cluster are listed in the last two columns of Table \ref{tab:Table1}.

We establish selection criteria for BSS and MS-TO stars following \citet{Raso2017} and \citet{Ferraro2018}. For a uniform selection criteria for all our clusters, we follow these steps:

\begin{enumerate}
\item We fit PARSEC isochrones of appropriate ages and metallicities to all the clusters. We then shift the isochrones  0.75 magnitude brighter on the main sequence to separate the binary stars from BSS and fit a zero-age main-sequence (PARSEC isochrone of 90--160 Myr age) to all the clusters.

\item According to the clusters' ages, we divide the 11 OCs into two categories: category I -- intermediate-age clusters with age < 6 Gyr, and category II -- old age clusters with age $\geq$ 6 Gyr. We use this step because intermediate-age clusters have a blue hook near the main-sequence turn-off as well as a larger range in the G magnitude above the turn-off point than the old age clusters.

\item We then normalize the CMDs, i.e., shift magnitude and colour of cluster members to locate the MS-TO point at (0,0) as shown in Figure \ref{fig:Figure 1}, where we represent shifted magnitude and color by $G^*$ and $(Bp-Rp)^*$, respectively.
\end{enumerate}

In order to select our MS-TO stars, we define a trapezoid around the MS-TO level as follows. In terms of magnitudes, we define an upper limit at $G^*$ = 0.0 (i.e. the TO), and a lower limit at $G^*$ = $-$1.1 for intermediate age cluster and $G^*$ = $-$0.75 for old clusters. In terms of colours, we define the redward limit at $(Bp-Rp)^*$ = 0.2. For the blueward limit, we choose a slanting line defined as, $G^*=-15\times(Bp-Rp)^*-1$, such that the known BSS populations with radial velocities available in the literature, are not lost as MS-TO stars. Our BSS candidates are sources that occupy the region between zero-age main-sequence, binary track and MS ridge line. To separate BSS candidates from the sources close to the binary track, we draw a red border at $(Bp-Rp)^*=-0.067$ for old clusters, and at 0.02 mag bluer than the corresponding ($Bp-Rp)^*$ value of the blue hook, for the intermediate age clusters. Finally, the sources along the SGB and RGB evolutionary track of PARSEC isochrone with $(Bp-RP)^*$ > 0.2 are identified as SGBs, RGBs, and RCs. The number of BSS candidates of the 11 OCs after applying the BSS selection criteria are listed in Table \ref{tab:Table1}. Figure \ref{fig:Figure 1} shows normalized CMDs of the 11 OCs with the MS-TO selection box, and BSS selection region, marked on individual CMDs. The BSS are shown as blue filled squares and the SGBs, RGBs, and RCs are shown as orange filled circles.

\section{Calculation of \texorpdfstring{$A^+$}{A+}}
\label{section:Calculation-of-A}
In order to measure the BSS sedimentation level in the OCs, we calculate $A^+,$ i.e., the area enclosed between the cumulative radial distribution of BSS and a reference population (hereafter REF). Depending on the completeness level up to $G_{\mathrm{TO}}$ of the clusters, we have used either SGBs, RGBs, and RCs or MS-TO stars, SGBs, RGBs, and RCs as REF in this work. The area enclosed between the cumulative radial distributions of BSS ($\phi_{\mathrm{BSS}}$) and REF ($\phi_{\mathrm{REF}}$) is given as:
\begin{equation} A^+ = \int^x_{x_{min}} \phi_{\mathrm{BSS}}(x^{\prime})- \phi_{\mathrm{REF}}(x^{\prime})dx^{\prime} 
\end{equation} 
where $x$ = $\mathrm{\log}(r/r_{\mathrm{h}})$ and $x_{\mathrm{min}}$ are the outermost and the innermost radii from the cluster center, respectively, and $r_{\mathrm{h}}$ is the half-mass radius of the cluster. Following \citet{Lanzoni2016}, we calculate $A^+$ only up to  one $r_{\mathrm{h}}$ of the cluster (hereafter $A^+_{\mathrm{rh}}$), mainly to compare the parameters in different stellar systems and take into account the cluster portion that is most sensitive to the phenomenon of BSS sedimentation that occurs due to DF. Moreover, to maximize the sensitivity of the parameter $A^+_{\mathrm{rh}}$, we plot the cumulative radial distributions as a function of the logarithm of the radius normalized to the cluster $r_{\mathrm{h}}$. To estimate $r_{\mathrm{h}}$, we first compute the projected half-light radius,  $r_{\mathrm{hp}}$ using equation 9 of \citet{Santos2020}, given as
$\mathrm{log}\left(\frac{r_{\mathrm{hp}}}{r_{\mathrm{c}}}\right) = -(0.339 \pm 0.009) + (0.602 \pm 0.015)c - (0.037 \pm 0.005)c^2$
where c = $\mathrm{log}(r_{\mathrm{t}}/r_{\mathrm{c}})$ is the concentration parameter, $r_{\mathrm{c}}$ is the core radius, and $r_{\mathrm{t}}$ is the tidal radius of the cluster obtained by fitting the King model \citep{King1962}. The readers are referred to the \S \ref{section:Trc} for details of the King model fitting. We then estimate 3D half-mass radius, $r_{\mathrm{h}}$, using the relation $r_{\mathrm{h}}$ = 1.33$ \times r_{\mathrm{hp}}$ \citep{Baumgardt2010}. The values of  $r_{\mathrm{h}}$ are listed in Table \ref{tab:Table2}. Figure \ref{fig:Figure 2} shows the cumulative radial distributions of the BSS and the REF population of the 11 OCs. The grey shaded portion enclosed between the BSS and the REF cumulative radial distributions corresponds to the estimated value of $A^+_{\mathrm{rh}}$ for the respective cluster and is noted in Table \ref{tab:Table2}. To estimate errors in $A^+_{\mathrm{rh}}$ values, we calculated $A^+_{\mathrm{rh}}$ for 1000 random samples of BSS and REF generated using the bootstrap method. We then determine the mean and dispersion of the 1000  $A^+_{\mathrm{rh}}$ values, which are within the uncertainties estimated by the bootstrap method. We consider the dispersion as an error in $A^+_{\mathrm{rh}}$ value of the cluster \citep{Lupton1993}. The errors in $A^+_{\mathrm{rh}}$ values are listed in Table \ref{tab:Table2}. The errors in $A^+_{\mathrm{rh}}$ are essentially dominated by the number of BSS in the clusters and their radial distribution with respect to the REF.

\section{Estimation of the Central relaxation time} 
\label{section:Trc}
\begin{figure}
	\includegraphics[width=0.35\textheight]{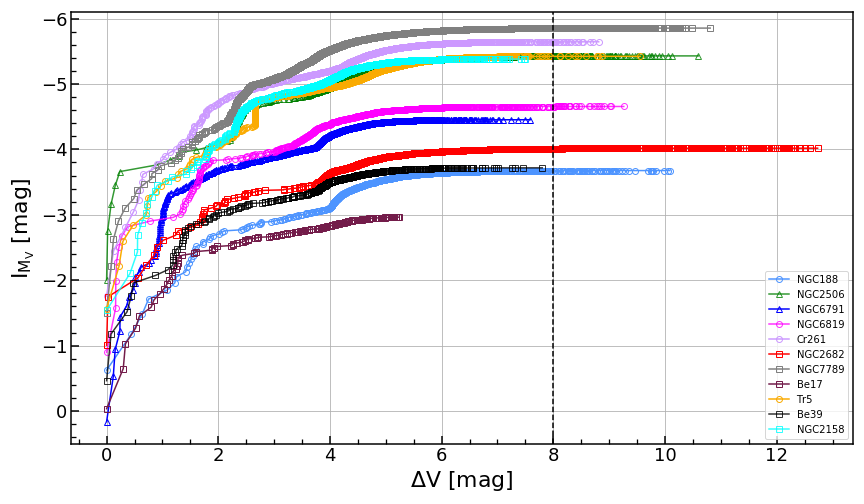}
    \caption{Integrated absolute magnitude profiles for the 11 OCs. $\Delta$V is the magnitude difference between each individual cluster member and the brightest cluster member. The black dashed line shows the saturation level of the integrated absolute magnitude profiles.} 
    	\label{fig:Figure 3}
\end{figure}
We used the same method as used by \citet{Ferraro2012}, \citet{Lanzoni2016} and \citet{Vaidya2020}, to calculate the $N_{\mathrm{relax}}$. For that, we first determine the central relaxation time, $t_{\mathrm{rc}}$, using the Equation 10 of \citet{Djorgovski1993}, $t_{\mathrm{rc}}$ = $1.491 \times 10^7 yr \times \frac{k}{ln(0.4N_{*})} <m_{*}>^{-1} \rho_{_{\scriptscriptstyle  M,O}}^{1/2} r^3_{c} $, where $k \sim 0.5592$, $r_{\mathrm{c}}$ is the core radius, $<m_{*}>$ is the average mass of the cluster members, $\rho_{_{\scriptscriptstyle  M,O}}$ is the central mass density of the cluster, and $N_*$ is the total number of the cluster members. We fitted the King model \citep{King1962} to the cluster members brighter than G = 17 mag, since the Gaia DR2 data is essentially complete down to G $\sim$ 17 mag. We divided the cluster radius into equal radius bins, and estimated surface density in each bin. We then plotted the logarithm of the estimated surface densities versus the logarithm of the radii and fitted the King model. We re-estimated the King model parameters for the 6 OCs taken from \citet{Vaidya2020}, using stars brighter than G = 17 mag. For all the clusters, the re-estimated $r_{\mathrm{c}}$ values are in agreement with the $r_{\mathrm{c}}$ values provided by \citet{Vaidya2020} except for NGC 188. In contrast, the $r_{\mathrm{t}}$ values are smaller than estimated by \citet{Vaidya2020}. Our estimated king model parameters are in close agreement with \citet{Kharchenko2013}. For Berkeley 17, we adopted these parameters, $r_{\mathrm{c}}$ and $r_{\mathrm{t}}$, from \citet{Bhattacharya2019}. The estimated values of $r_{\mathrm{c}}$, $r_{\mathrm{t}}$, and c of the 11 OCs are listed in Table \ref{tab:Table2} and Figure \ref{fig:Figure A6} shows only the newly fitted King profiles of the four clusters in this work, Collinder 261, NGC 2682, NGC 7789, and Trumpler 5.

To estimate $N_*$ first, we converted the G magnitudes into V magnitudes using the relation available on the Gaia-ESO website\footnote{\url{https://gea.esac.esa.int/archive/documentation/GDR2/Data_processing/}} and then computed apparent integrated magnitudes of the clusters using the following correlation, as illustrated by \citet{Piskunov2008}.
\begin{equation*}
I_{\mathrm{V}} = -2.5 \mathrm{log} \left( \sum_i^{N_i} 10^{-0.4 V_i} + 10^{-0.4 \Delta I_{\mathrm{V}}} \right)
\end{equation*}
where $N_i$ and $V_i$ are the number and the apparent V magnitude of the cluster members. $\Delta I_{\mathrm{V}}$ is the term proposed to perform unseen stars correction, i.e., to make $I_{\mathrm{V}}$ and $I_{M_{\mathrm{V}}}$ independent of the extent of the stellar magnitudes observed in a cluster. We then converted the integrated apparent magnitudes into integrated absolute magnitudes by using the distance modulus and extinction of the clusters (distances of the 11 OCs are listed in Table \ref{tab:Table1}). Figure \ref{fig:Figure 3} shows the integrated absolute magnitude profiles of the 11 OCs, where $\Delta V$ is the magnitude difference between each individual cluster member and the brightest cluster member. All the cluster profiles are seen to saturate at $\Delta V$ = 8 mag except for Berkeley 17, which is deficient in stars fainter than $\Delta V$ $\sim$5.2 mag. This implies that beyond $\Delta V$ = 8 mag, the contribution from the faint cluster members to the integrated absolute magnitudes of the clusters is negligible. In order to perform unseen stars correction in Berkeley 17, we used NGC 2506 as a template cluster, and estimated the difference in its integrated absolute magnitudes at the faintest magnitude of Berkeley 17 and the brightest magnitude of Berkeley 17 plus $\Delta V$ = 8 mag. This  correction term, 0.00156 mag, was subtracted from the integrated absolute magnitude of Berkeley 17. \citet{Lata2002} also estimated integrated absolute magnitudes in the V band for Berkeley 39, NGC 188, NGC 2158, NGC 2506, NGC 2682, NGC 6791, and Trumpler 5. Our estimated values of integrated absolute magnitudes are slightly larger than those estimated by \citet{Lata2002} for all the clusters except NGC 2158, for which our estimation is larger by a factor of 2.

We then converted the integrated absolute magnitudes of all the clusters into luminosities. \citet{Piskunov2011} provide the log(M/L$_{\mathrm{V}}$) versus age relation for open clusters for a sample of 650 OCs. We used these luminosities to estimate the average masses of the clusters by averaging the masses corresponding to the lower and the upper bounds in the \citet{Piskunov2011} correlation between log(M/L$_{\mathrm{V}}$) and log(age) > 9. We then divide the total cluster mass by the average mass of cluster members, taken as the average mass of the main-sequence stars of the cluster, to get the $N_*$. Finally, to estimate the central mass density, we first estimate central luminosity density using the equation 6 of \citet{Djorgovski1993} $\rho = \mathrm{L_{V}}/\mu r_{\mathrm{c}}^2$, where $\mathrm{L_{V}}$ is the total luminosity of the cluster, and $\mu$ is a function which depends on the concentration (c) of the cluster, given as 
$\mathrm{log}\mu = -0.14192c^4 + 1.15592c^3 - 3.16183c^2 +4.21004c - 1.00951$
Using this central luminosity density, the central mass density of the clusters were estimated by taking an average of the lower and upper bound of log(M/L$_{\mathrm{V}}$) from \citet{Piskunov2011}. The estimated values of Integrated absolute magnitudes, central luminosity densities, central mass densities, and average stellar masses of the 11 OCs are listed in Table \ref{tab:Table2}. \citet{Heinke2020} estimated central mass densities of Collinder 261 as 15 $M_{\sun}$/pc$^3$, NGC 188 as 9$\pm$2 $M_{\sun}$/pc$^3$, NGC 2682 as 32 $M_{\sun}$/pc$^3$, NGC 6791 as 11 $M_{\sun}$/pc$^3$, and NGC 6819 as 54 $M_{\sun}$/pc$^3$. For NGC 188, we obtained the similar value of central mass density to that of \citet{Heinke2020}. In contrast, our estimations are smaller for NGC 2682 and NGC 6819, whereas larger for Collinder 261 and NGC 6791.

\section{Results and Discussion} 
\label{section:Results}
\begin{table*}
	\caption{The COCOR tool results of OCs and the less evolved GCs for the correlation between $A^+_{\mathrm{rh}}$ and structural and dynamical parameters of the clusters. Column 1 gives the fitted correlation, column 2 gives the name of the correlation coefficient where P refers to the Pearson correlation coefficient and S refers to the Spearman rank correlation coefficient, columns 3 and 4 list the calculated correlation coefficients for the OCs and the less evolved GCs, column 5 and 6 give the results obtained by employing the statistical tests to compare the correlation coefficients of the OCs and the less evolved GCs using the COCOR tool, the last column denotes whether the null hypothesis is or is not rejected, with a $\checkmark$ sign implying that the null hypothesis is not rejected.}
	\label{tab:Table3}
	\begin{tabular}{ccccccc}
		\hline
		\\
		Correlation  & \begin{tabular}{c} Correlation \\ coefficient \end{tabular} & OCs & GCs & \begin{tabular}{c} p-value \\ (Fisher test) \end{tabular} & \begin{tabular}{c} 95$\%$ CI$^{\star}$ \\ (Zou test) \end{tabular} & COCOR tool result \\
		 \\
		\hline
		\\
		  $A^+_{\mathrm{rh}}$ vs $N_{\mathrm{relax}}$ & \begin{tabular}{c} P \\ S \end{tabular} & \begin{tabular}{c} $+$  0.532 \\ $+$  0.629 \end{tabular} & \begin{tabular}{c} $+$0.768 \\ $+$0.763 \end{tabular} & \begin{tabular}{c}   0.2743 \\   0.4952 \end{tabular} &  \begin{tabular}{c}   $-$0.8754 -- $+$0.1262 \\   $-$0.7249 -- $+$0.1742 \end{tabular} &  \begin{tabular}{c} $\checkmark$ \\ $\checkmark$ \end{tabular} \\
		\\
		$A^+_{\mathrm{rh}}$ vs $r_{\mathrm{c}}$ & \begin{tabular}{c} P \\ S \end{tabular} & \begin{tabular}{c} $-$  0.570 \\ $-$  0.709  \end{tabular} & \begin{tabular}{c} $-$0.488 \\ $-$0.452 \end{tabular} &  \begin{tabular}{c}   0.7787 \\   0.3692 \end{tabular} &  \begin{tabular}{c}   $-$0.6522 -- $+$0.5911 \\   $-$0.7885 -- $+$0.3498 \end{tabular} &  \begin{tabular}{c} $\checkmark$ \\ $\checkmark$ \end{tabular} \\
		\\
		$A^+_{\mathrm{rh}}$ vs $r_{\mathrm{c}}/r_{\mathrm{e}}$ & \begin{tabular}{c} P \\ S \end{tabular} & \begin{tabular}{c} $-$  0.061 \\ $-$  0.209 \end{tabular} & \begin{tabular}{c} $-$0.544   \\ $-$0.467 \end{tabular} & \begin{tabular}{c}   0.4176 \\   0.6640 \end{tabular} &  \begin{tabular}{c} $-$  0.6913 -- $+$1.2192 \\   $-$0.8839 -- $+$1.0592 \end{tabular} &  \begin{tabular}{c} $\checkmark$ \\ $\checkmark$ \end{tabular}\\
		\\
		$A^+_{\mathrm{rh}}$ vs $c$ & \begin{tabular}{c} P \\ S \end{tabular} & \begin{tabular}{c} $+$  0.041 \\ $+$  0.082  \end{tabular} & \begin{tabular}{c} $+$0.428 \\ 0.325 \end{tabular} & \begin{tabular}{c}   0.3474 \\   0.5650 \end{tabular} &  \begin{tabular}{c}   $-$1.0820 -- $+$0.3785 \\   $-$-0.9717 -- $+$0.5199 \end{tabular} &  \begin{tabular}{c} $\checkmark$ \\ $\checkmark$ \end{tabular}\\
		\\
		\hline
	\end{tabular}
	\begin{tablenotes}
		\item {$^{\star}$Confidence interval}
	\end{tablenotes}
\end{table*}
\begin{figure}
	\includegraphics[width=0.35\textheight]{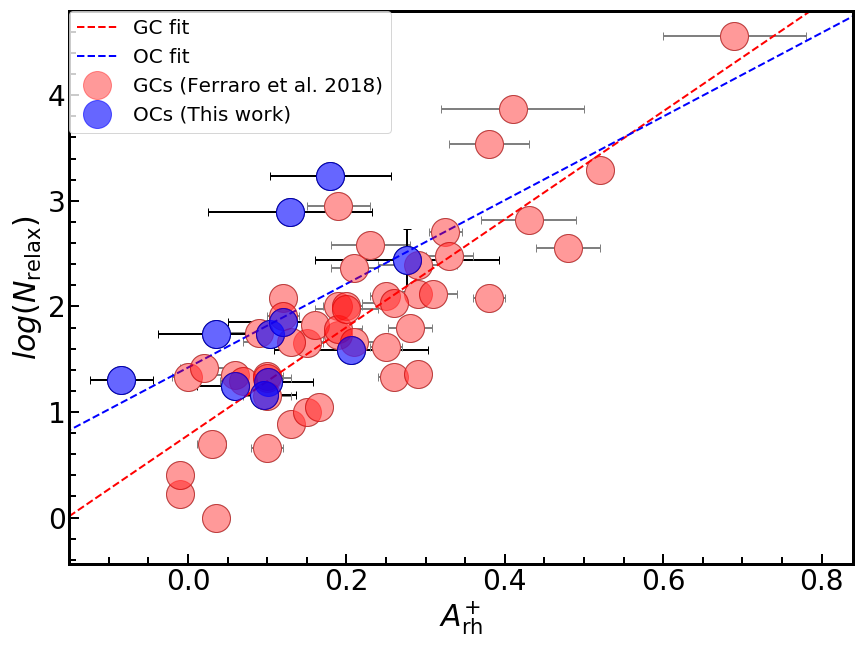}
    \caption{The correlation between the values of $A^+_{\mathrm{rh}}$ and the number of current central relaxation a cluster has undergone since its formation, $N_{\mathrm{relax}}$, for 11 OCs (blue filled circles) and 48 GCs (red filled circles) of \citet{Ferraro2018}. The blue dashed line represents the best fitted line for the OCs whereas the red dashed line shows the best-fit correlation \citep{Ferraro2018} for the GCs.} 
    	\label{fig:Figure 4}
\end{figure}
\begin{figure}
	\includegraphics[width=0.48\textwidth]{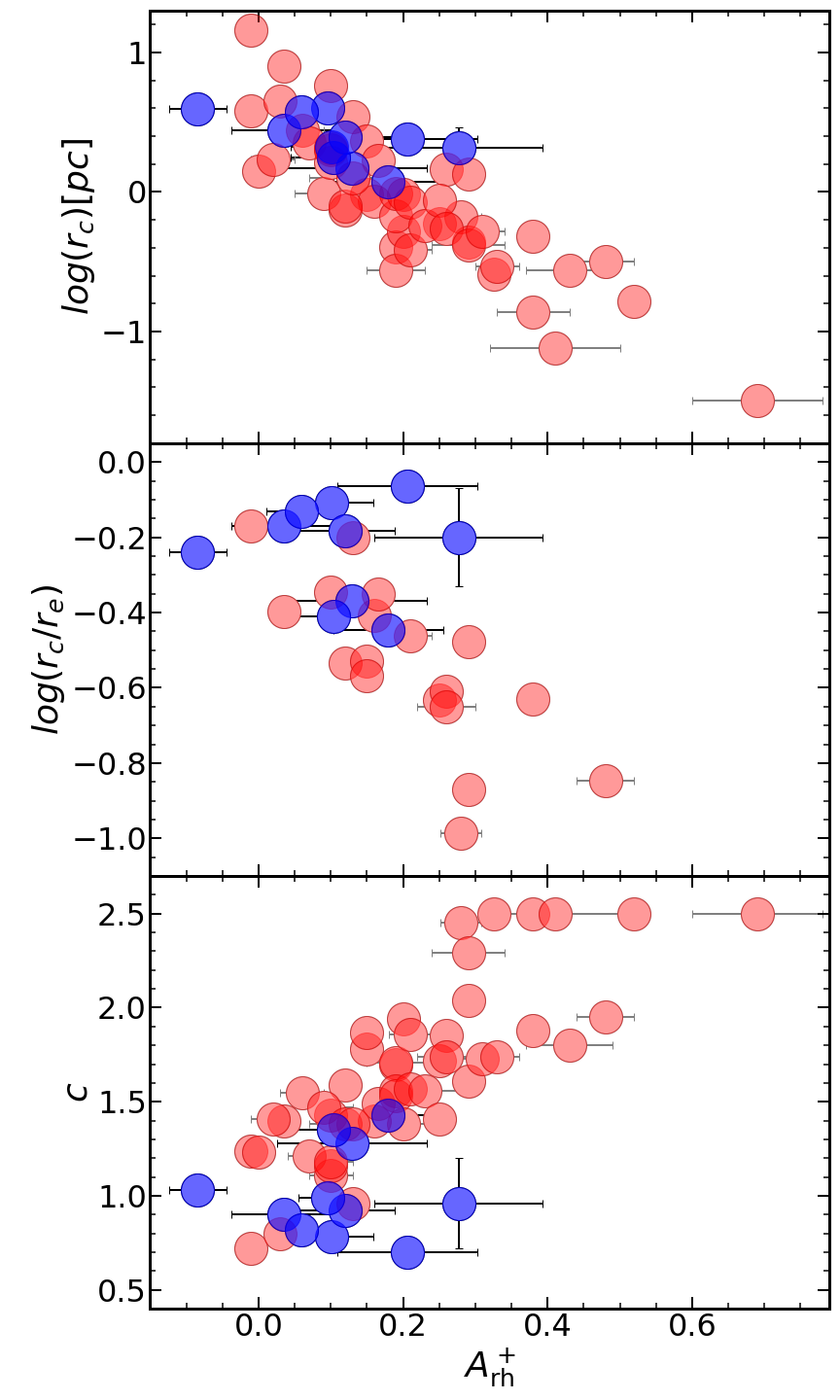}
    \caption{The correlation between $A^+_{\mathrm{rh}}$ and structural parameters for 11 OCs (blue filled circles) and GCs (red filled circles). We utilized the values of $A^+_{\mathrm{rh}}$ and structural parameters of GCs from \citet{Lanzoni2016} \citet{Ferraro2018}, and the references therein.}
    	\label{fig:Figure 5}
\end{figure}
\citet{Lanzoni2016}, \citet{Ferraro2018,Ferraro2020} ascertained that $A^+_{\mathrm{rh}}$ is a powerful indicator of the  dynamical evolution of GCs. We plot $A^+_{\mathrm{rh}}$ against $N_{\mathrm{relax}}$ for 11 OCs in Figure \ref{fig:Figure 4} to investigate whether  $A^+_{\mathrm{rh}}$ values of OCs truly indicate the dynamical status of the clusters. For comparison, we also show the two identical quantities of GCs from \citet{Ferraro2018} on the same figure. The best-fit relation for OCs plotted in Figure \ref{fig:Figure 4} is

\begin{equation}
	\mathrm{log}(N_{\mathrm{relax}}) = 4.0(\pm 2.1)\times A^{+}_{\mathrm{rh}}+1.42(\pm 0.30)
\end{equation}

Whereas the best-fit correlation for GCs is
\begin{equation}
	\mathrm{log}(N_{\mathrm{relax}}) = 5.1(\pm 0.5)\times A^{+}_{\mathrm{rh}}+0.79 (\pm 0.12)
\end{equation} 
The OC datapoints exhibit a positive correlation between $A^+_{\mathrm{rh}}$ and $ N_{\mathrm{relax}}$ with a smaller slope and a higher intercept than the correlation among the GC datapoints \citep{Ferraro2018}. Moreover, the OC correlation has a low statistical significance with the Spearman rank correlation coefficient being 0.629, and the Pearson correlation coefficient being 0.532. The OCs are seen to lie near the middle to the lower end of the plot, implying, as expectantly, that they are among the least dynamically evolved GCs. Moreover, physically it is not possible for OCs to be as evolved as GCs. Therefore we compare OCs with the less evolved GCs. In particular, we calculate the Spearman rank correlation coefficient, and the Pearson correlation coefficient for GCs with $\mathrm{log}(N_{\mathrm{relax}})$ $\le$ 3.38, and found them to be 0.763 and 0.768, respectively. Since the sample sizes of OCs and GCs are different, a direct comparison between the correlation coefficients is not sensible. Therefore, we use the COCOR\footnote{\url{http://comparingcorrelations.org}} tool, which covers a broad range of tests including the comparisons of independent and dependent correlations with either overlapping or nonoverlapping variables \citep{diedenhofen2015}. The COCOR tool, however, is limited to the comparison between two correlations only. In our case, the correlation coefficients are independent to each other, so it compares the correlation coefficients through two tests: the Fisher test \citep{fisher1992}, and the Zou test \citep{zou2007}. We chose the values of $\alpha$, p-value threshold, and the Zou's confidence level as 0.05 and 0.95, respectively. We then conduct a two-tailed test that gives the results in the form of whether the two correlations are equal or not. We get the p-value, i.e., the significance level of the comparison done by the Fisher test as 0.2743 and 0.4952 for the Pearson correlation coefficient and the Spearman rank correlation coefficient, respectively. We get the Zou's confidence interval as $-$0.8754 -- $+$0.1262 and $-$0.7249 -- $+$0.1742 for Pearson correlation coefficients and  Spearman rank correlation coefficients, respectively. The p-value estimated by the Fisher's test is greater than the chosen $\alpha$ value and the estimated Zou's confidence intervals also contain zero which indicate that the null hypothesis that the two distributions are similar is not rejected. Hence, these tests suggest that the correlation coefficients estimated for the less evolved GCs and OCs are not different. However, given the small sample size of OCs, we need a larger sample size to see if they are similar.
\begin{figure}
	\includegraphics[width=0.48\textwidth]{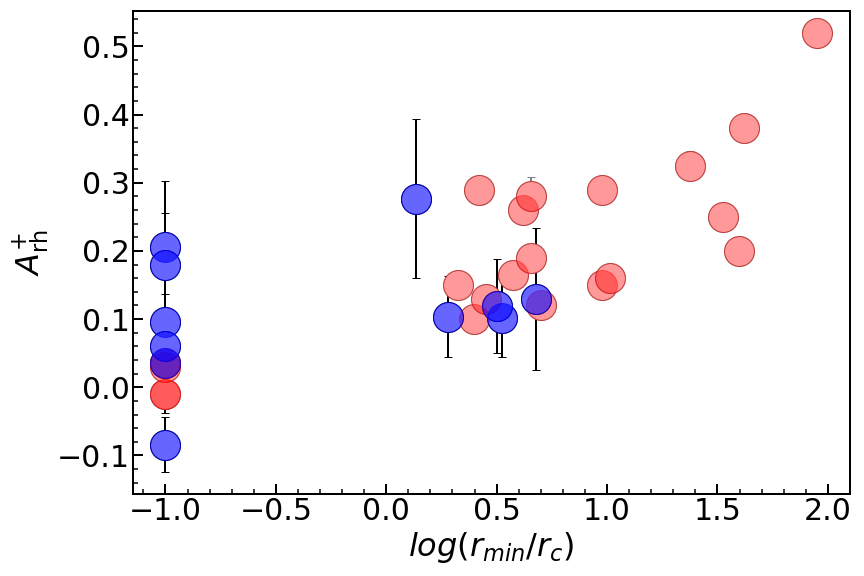}
    \caption{The correlation between $A^+_{\mathrm{rh}}$ and the location of minima in BSS radial distributions, $\mathrm{log}(r_{\mathrm{min}}/r_{\mathrm{c}})$, for OCs (blue filled circles) and GCs (red filled circles) of \citet{Lanzoni2016}. Following \citet{Lanzoni2016}, we consider the value of $r_{\mathrm{min}}/r_{\mathrm{c}}$ as 0.1 for those OCs which have flat BSS radial distributions.}
    	\label{fig:Figure 6}
\end{figure}
\citet{Ferraro2018,Ferraro2020} demonstrated that the long-term dynamical evolution tends to produce compact stellar systems. To show this effect in OCs, we plot $A^+_{\mathrm{rh}}$ against three structural parameters, $r_{\mathrm{c}}$, $c$, and $r_{\mathrm{c}}/r_{\mathrm{e}}$ (the ratio of core radius to effective radius of the cluster) as shown in Figure \ref{fig:Figure 5}. For this, we derived the effective radius, defined as the radius of the circle in projection including half of the total counted stars of a cluster for all the 11 OCs included in the present work. The values of distances, and the ages of $6$ clusters are utilized from \citep{Vaidya2020}, while for Berkeley 17, we use these parameters derived by \citet{Bhattacharya2019} and for the remaining 4 OCs, we determine the ages and the distances by fitting the PARSEC isochrones to the cluster members. The values of the ages and distances of the clusters are listed in Table \ref{tab:Table1}. In Figure \ref{fig:Figure 5}, we plot our estimated values of $A^+_{\mathrm{rh}}$ against these structural parameters, $\mathrm{log}(r_{\mathrm{c}})$, $\mathrm{log}(r_{\mathrm{c}}/r_{\mathrm{e}})$, and $c,$ in the top, middle, and the bottom panel, respectively, for 11 OCs. As can be seen in Figure  \ref{fig:Figure 5}, OCs follow the same trend as GCs, and occupy the same parameter space with smaller values of $A^+_{\mathrm{rh}}$. We calculated the Pearson correlation coefficients and the Spearman rank correlation coefficients for OCs of all the three plots of Figure \ref{fig:Figure 5}. In order to compare OCs with GCs, we also calculate the correlation coefficients for less evolved GCs, i.e., GCs which are having values $\mathrm{log}(r_{\mathrm{c}})$, $\mathrm{log}(r_{\mathrm{c}}/r_{\mathrm{e}})$, and $c$ similar as OCs. The estimated correlation coefficients of OCs and GCs for all the three cases and the COCOR tool results are listed in the Table \ref{tab:Table3}. Again, we see that they are not different within the errors, but we need a larger sample size of OCs since errors are large due to our small sample size.

Since $A^+_{\mathrm{rh}}$ and $r_{\mathrm{min}}$ both have been used as indicators of the dynamical ages of the clusters, it is useful to see how these two observed parameters of clusters themselves correlate with one another. In fact, \citet{Lanzoni2016} plotted $A^+_{\mathrm{rh}}$ against the logarithm of $r_{\mathrm{min}}$ (in the units of $r_{\mathrm{c}}$) for their GC sample and found that these two quantities are in a linear correlation with each other. We plot these two parameters against one another for our 11 OCs (blue filled circles) and show them in Figure \ref{fig:Figure 6}. We use the values of $\mathrm{log}(r_{\mathrm{min}})$ of 4 OCs with bimodal BSS radial distributions from \citet{Vaidya2020} whereas for Berkeley 17 we use the value derived by \citet{Bhattacharya2019}. Collinder 261 \citep{Rain2020a} and Trumpler 5 \citep{Rain2020b} and two OCs in \citet{Vaidya2020} have flat BSS radial distributions. For NGC 7789 and NGC 2682, we plotted BSS radial distributions using the method described by \cite{Vaidya2020}. As shown in Figure \ref{fig:Figure A6}, these two clusters also have flat BSS radial distributions. Thus we have 5 OCs with bimodal distributions and 6 OCs with flat BSS radial distributions for which we consider $r_{\mathrm{min}}/r_{\mathrm{c}}$ as 0.1. For a comparison, we also plot all the GCs data points (red filled circles) from \citet[their Figure 4]{Lanzoni2016}. The OCs parameters, $A^+_{\mathrm{rh}}$ and $\mathrm{log}(r_{\mathrm{min}}/r_{\mathrm{c}})$, appear to occupy the same parameter space of GCs. Not surprisingly, the OCs are among the least dynamically evolved GCs.  

Among the 11 OCs studied in the current work, Trumpler 5 is the least dynamically evolved. Its value of $A^+_{\mathrm{rh}}$ is also consistent with the value of $N_{\mathrm{relax}}$ (see Figure \ref{fig:Figure 4}). \citet{Rain2020b} presented the BSS radial distribution of Trumpler 5 using RGBs as a reference population and found it to have flat BSS radial distribution, which is in agreement with our finding that the cluster is dynamically young. The values of $N_{\mathrm{relax}}$ and $A^+_{\mathrm{rh}}$ of  NGC 7789 and Collinder 261 are consistent with each other, which shows that the clusters are dynamically intermediate-age. In contrast to our conclusion of the dynamical status of NGC 7789, \citet{Wu2007} reported that the cluster is mass-segregated by estimating concentration parameter for different mass range sources and by fitting mass-function in different spatial ranges. However, neither our estimate of the number of relaxations undergone by the cluster since its formation nor our estimated $A^+_{\mathrm{rh}}$ suggests that NGC 7789 is a dynamically evolved cluster. \citet{Rain2020a} presented the BSS radial distribution of Collinder 261 using RGBs as a reference population and found it to have flat BSS radial distribution. We notice that NGC 2682 is a cluster that is expected to be a dynamically evolved cluster according to previous works in the literature in which signatures of extra-tidal sources and mass segregation have been found \citep{Fan1996,Bonatto2003,Geller2015,Carrera2019}. According to our analysis, its $A^+_{\mathrm{rh}}$ and $N_{\mathrm{relax}}$ suggest that it is of intermediate dynamical age. \citet{Vaidya2020} had shown the evidence of the dynamical ages of 7 OCs using BSS radial distributions. Among those, 5 OCs, Melotte 66, NGC 188, NGC 2158, NGC 2506, and NGC 6791, were classified as intermediate dynamical age clusters, and 4 of them studied in the current work are consistent with the present finding. Berkeley 39 and NGC 6819 were found to show flat radial distribution \citep{Vaidya2020}. However, as per the $A^+_{\mathrm{rh}}$ and $N_{\mathrm{relax}}$ values, our current work finds these clusters to be of intermediate dynamical age.

\section{Summary} 
\label{section:Summary}

The present study is the extended version of the work done by \citet{Vaidya2020}, in which they identified BSS and RGBs of 7 OCs and performed the analysis of dynamical evolution of the clusters using the BSS radial distributions. This work presents the first ever attempt at estimating $A^+_{\mathrm{rh}}$ in 11 OCs, including the previous 6 clusters studied by \citet{Vaidya2020}. While the correlation between $A^+_{\mathrm{rh}}$ and  $r_{\mathrm{min}}$ is not clear for OC datapoints alone, the use of  $A^+_{\mathrm{rh}}$ as a tracer of dynamical evolution is clear from its relation to $N_{\mathrm{relax}}$ (Figure \ref{fig:Figure 4}). Our study shows that the $A^+_{\mathrm{rh}}$ when plotted against the theoretical estimation of the relaxation status of the clusters, $N_{\mathrm{relax}}$, or against the structural parameters of the clusters, the OC datapoints are seen to fall in the category of the less evolved GCs. OCs are young stellar systems of $10^6$ $-$ $10^9$ Gyr age \citep{Lada2010}, in contrast with GCs formed during the early stages of the Milky Way \citep{Vandenberg1996}. Also, GCs have a dense star distribution with  $\sim10^5$ member stars while OCs are sparser, containing only $\sim10^3$ member stars in general. Because of the high density, GCs have enough gravitational force to resist the tidal force and remain in a spherical shape and gravitationally bounded \citep{Harris1979,Freeman1981} while sparser OCs are more easily stretched by the external forces and do not remain gravitationally bound over time and spread out \citep{Chen2004,Zhai2017,Bhattacharya2017,Bhattacharya2021}. Therefore, we compared the correlation coefficients of OCs with the less evolved GCs using the statistical tests, the Fisher test and the Zou test, implemented by the COCOR tool. The employed statistical tests suggest that the compared correlation coefficients are consistent within the large errors. In order to determine a more sound correlation between the $A^+$ and the other markers of the dynamical ages of the OCs, we will extend this work using the recently released Gaia EDR3 data through an application of an automated membership determination algorithm \citep[ML-MOC;][]{Agarwal2020} to a large number of OCs having greater than 10 -- 12 BSS.

\section*{Acknowledgements}

We thank the anonymous referee for their valuable comments. This work has made use of second data release from the European Space Agency (ESA) mission {\it Gaia} (\url{https://www.cosmos.esa.int/gaia}), Gaia-DR2 \citep{Gaia2018}, processed by the {\it Gaia} Data Processing and Analysis Consortium (DPAC, \url{https://www.cosmos.esa.int/web/gaia/dpac/consortium}). Funding for the DPAC has been provided by national institutions, in particular the institutions participating in the {\it Gaia} Multilateral Agreement. This research has made use of the VizieR catalog access tool, CDS, Strasbourg, France. This research made use of {\small ASTROPY}, a {\small PYTHON} package for astronomy \citep{Astropy2013}, {\small NUMPY} \citep{Harris2020}, {\small MATPLOTLIB} \citep{Hunter4160265}, and {\small SCIPY} \citep{Virtanen2020}. This research also made use Astrophysics Data System (ADS) governed by NASA (\url{https://ui.adsabs.harvard.edu}).

\section*{Data availability}
The data underlying this article are publicly available at \url{https://gea.esac.esa.int/archive}. The derived data generated in this research will be shared on reasonable request to the corresponding author.




\bibliographystyle{mnras}
\bibliography{References} 

\begin{landscape}
\begin{table}
	\caption{The dynamical and the structural parameters, and the estimated values of $A^+_{\mathrm{rh}}$ and errors in $A^+_{\mathrm{rh}}$ of the OCs. Here, Column 1: Cluster name; Column 2-4, 6, and 7: Fitted King parameters; Column 5: Minima in radial distribution of BSS relative to cluster $r_{\mathrm{c}}$; Column 8: Integrated absolute magnitudes; Column 9 and 10: central luminosity and mass density, respectively; Column 10: average stellar mass; Column 11 and 12: Central relaxation time and Number of central relaxations experienced by a cluster since its formation, respectively.}
	\label{tab:Table2}
	\begin{tabular}{ccccccccccccccc}
		\hline
		\\
	    Cluster  & $r_{\mathrm{c}}$ & $r_{\mathrm{t}}$ & $r_{\mathrm{h}}$ & $r_{\mathrm{min}}/r_{\mathrm{c}}$ & $r_{\mathrm{e}}$ & c & $I_{\mathrm{M_V}}$ & $\rho_{_{\scriptscriptstyle  L,O}}$ & $\rho_{_{\scriptscriptstyle  M,O}}$ & <m$_*$> & $t_{\mathrm{rc}}$ &$N_{\mathrm{relax}}$ &$A^+_{\mathrm{rh}}$ & Error  \\
	      & (arcmin) & (arcmin) & (arcmin) &  & (arcmin) &  &  mag &  $L_{\sun}$/$pc^3$ &  $M_{\sun}$/$pc^3$ &  $M_{\sun}$ & (Myr) & & & ($\epsilon_{A^+}$)  \\
		 \\
		\hline
		\\
 Berkeley 17$^a$  & 2.28$\pm$0.8 & 21$\pm$9 & 4.90$\pm$1.40 & 1.37 &  3.70$\pm$1.10 & 0.96$\pm$0.24 & $-$ 2.97 &  10.0$\pm$8.0 &  7.0$\pm$5.0  &  0.91  &  36$\pm$24  &  280$\pm$180  & 0.289 & 0.123\\
 Berkeley 39$^b$  & 2.24$\pm$0.19 & 17.72$\pm$1.51 & 4.43$\pm$0.26 & -- &  3.33$\pm$0.20  & 0.90$\pm$0.05 &  $-$3.72  &  24.0$\pm$5.0  &  15.8$\pm$2.9  &  1.00  &   109$\pm$18  &  55.0$\pm$9.0 & 0.036 & 0.072 \\ 
 Collinder 261$^b$  & 2.83$\pm$0.18 & 27.44$\pm$3.58 & 6.2$\pm$0.5 & -- &  4.69$\pm$0.35  & 0.99$\pm$0.06 &  $-$5.64  &  60$\pm$10  &  39$\pm$7  &  0.97  &  410$\pm$50  &  14.6$\pm$1.9  & 0.096 & 0.041\\
 NGC 188$^b$ & 2.80 $\pm$0.12 & 53.30$\pm$7.62 & 8.7$\pm$0.7 & 4.76$*$ &  6.60$\pm$0.50  & 1.28$\pm$0.06 &  $-$3.68  &  6.4$\pm$0.9  &  4.1$\pm$0.6  &  0.92  &  8.2$\pm$0.7  &  780$\pm$70  & 0.128 & 0.108 \\ 
 NGC 2158$^b$  & 1.69$\pm$0.11 & 10.25$\pm$0.21 & 2.9$\pm$0.09 & 3.33$*$ &  2.18$\pm$0.07  & 0.78$\pm$0.03 &  $-$5.38  &  340$\pm$40  &  222$\pm$28  &  1.29  &  114$\pm$15  &  19.3$\pm$2.6  & 0.101 & 0.058 \\ 
 NGC 2506$^b$  & 1.94$\pm$0.07 & 43.40$\pm$8.10 & 6.6$\pm$0.6 & 1.90$*$ &  4.90$\pm$0.50  & 1.35$\pm$0.08 &  $-$5.43  &  87$\pm$13  &  56$\pm$8  &  1.06  &  40.2$\pm$3.5  &  55$\pm$5  & 0.104 & 0.060 \\ 
 NGC 2682$^b$  & 4.78$\pm$0.12 & 129.91$\pm$17.31 & 17.8$\pm$1.2 & -- &  13.4$\pm$0.90  & 1.43$\pm$0.06 &  $-$4.02  &  1.43$\pm$0.14  &  0.93$\pm$0.09  &  0.82  &  2.32$\pm$0.14  &  1720$\pm$100  & 0.180 & 0.077  \\ 
 NGC 6791$^b$  & 1.90$\pm$0.14 & 15.77$\pm$1.74 & 3.85$\pm$0.26 & 3.16$*$ &  2.90$\pm$0.20  & 0.92$\pm$0.06 &  $-$4.45  &  76$\pm$14  &  49$\pm$9  &  1.00  &  124$\pm$19  &  70$\pm$11  & 0.119 & 0.067  \\ 
 NGC 6819$^b$  & 3.12$\pm$0.19 & 15.62$\pm$0.43 & 3.71$\pm$0.12 & -- &  3.61$\pm$0.11  & 0.7$\pm$0.03 &  $-$4.66  &  36$\pm$4  &  22.9$\pm$2.5  &  1.08  &  71$\pm$9  &  39$\pm$5  & 0.206 & 0.095  \\
 NGC 7789$^b$  & 6.58$\pm$0.30 & 43.64$\pm$3.84 & 11.8$\pm$0.6 & -- &  8.90$\pm$0.50  & 0.82$\pm$0.04 &  $-$5.85  &  8.2$\pm$1.1  &  5.3$\pm$0.7  &  1.15  &  107$\pm$11  &  17.8$\pm$1.8  & 0.06 & 0.048 \\
 Trumpler 5$^b$  & 4.19$\pm$0.17 & 45.21$\pm$7.99 & 9.8$\pm$0.9 & -- &  7.30$\pm$0.70  & 1.03$\pm$0.08 &  $-$5.42  &  13.9$\pm$2.2  &  9.0$\pm$1.5  &  1.09  &  175$\pm$18  &  20.0$\pm$2.0  & -0.084 & 0.040 \\ 
		\\
		\hline
	\end{tabular}
	\begin{tablenotes}
		\item {The values of $r_{\mathrm{c}}$ and $r_{\mathrm{t}}$ are taken from: $^a$\citet{Bhattacharya2019}; $^b$This work }
		\item {The values of $r_{\mathrm{min}}$  are taken from: $^a$\citet{Bhattacharya2019}; $^*$\citet{Vaidya2020}}
	\end{tablenotes}
\end{table}
\end{landscape}

\appendix
\section{ Membership determination and MS-TO stars identification in Collinder 261, NGC 2681, NGC 7789, and Trumpler 5} 
\label{Appendix}
\begin{figure*}
	\begin{subfigure}[b]{1.0\textwidth}
		\includegraphics[width=1.0\textwidth]{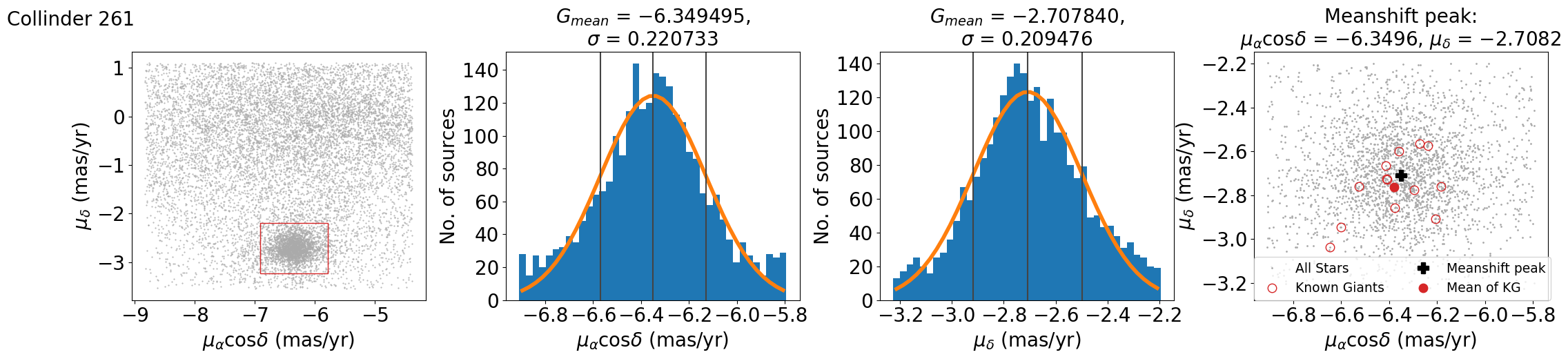}
		\caption*{}
	\end{subfigure}
	\quad
	\begin{subfigure}[b]{1.0\textwidth}
		\includegraphics[width=1.0\textwidth]{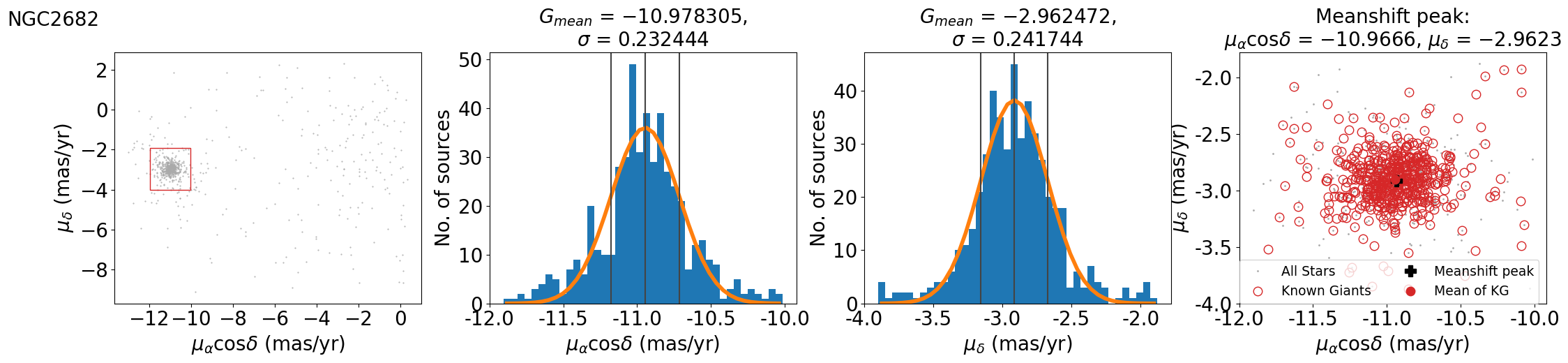}
		\caption*{}
	\end{subfigure}
	\quad
	\begin{subfigure}[b]{1.0\textwidth}
		\includegraphics[width=1.0\textwidth]{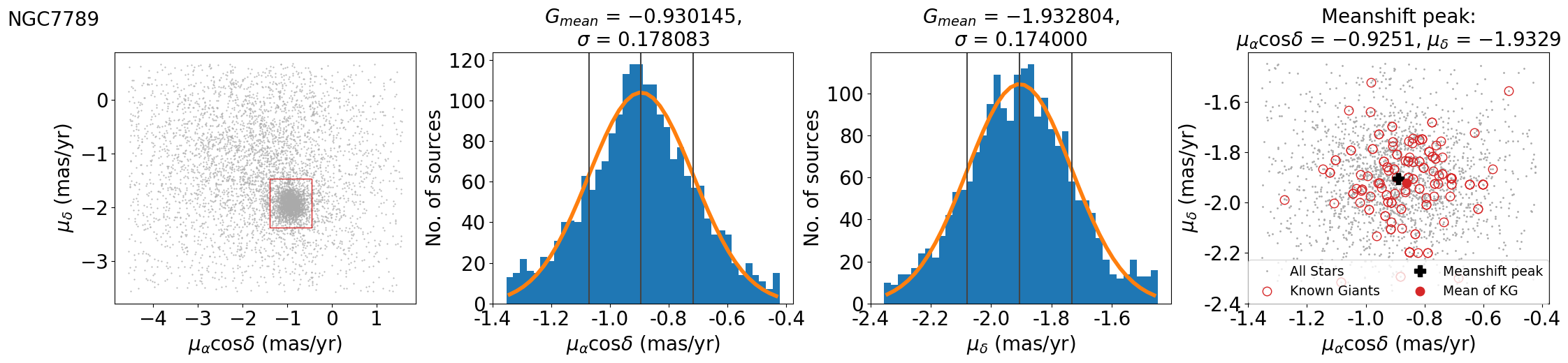}
		\caption*{}
	\end{subfigure}
	\quad
	\begin{subfigure}[b]{1.0\textwidth}
		\includegraphics[width=1.0\textwidth]{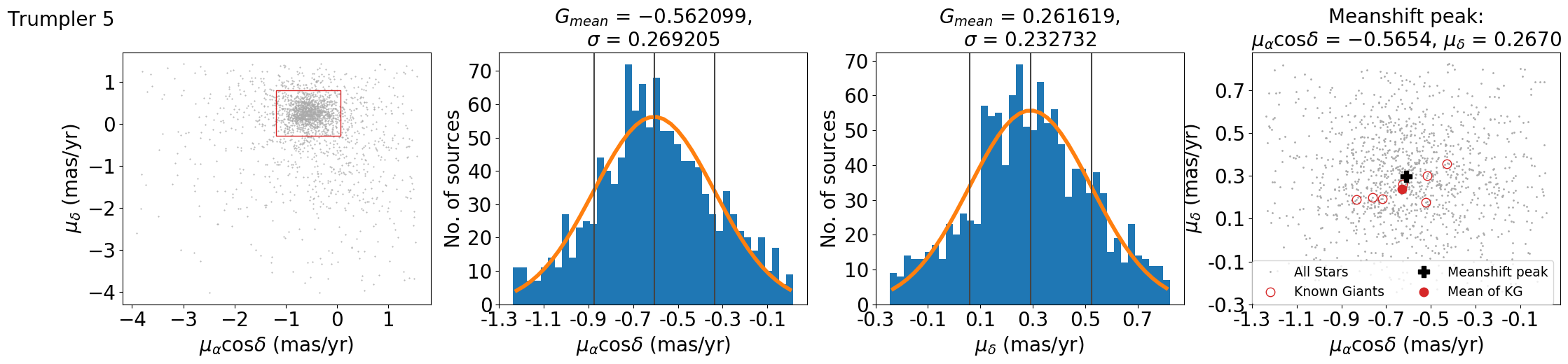}
		\caption*{}
	\end{subfigure}
	\caption{The proper motion selection criteria of the cluster members. The left panels show the proper motion scatter diagram of all sources within $10^{\prime}$ radius of the cluster centers. The red rectangular region shows our initial range of proper motion selected by visual examination of the proper motion scatter diagrams. The two middle panels show the frequency distribution of proper motions in RA and DEC, respectively, with orange curve showing the fitted Gaussian function to the distributions. The proper motion scatter-diagram of the cluster members selected within $G_{mean}$ $\pm$ 2.5$\sigma$ range are shown in the rightmost panels, where the black plus sign shows the mean of those sources computed using the mean-shift algorithm, whereas the red open circles and the filled circle are spectroscopically confirmed cluster members from the literature and their mean, respectively.}
	\label{fig:Figure A1}
\end{figure*}

\begin{figure*}
	\begin{subfigure}[b]{0.45\textwidth}
    		\includegraphics[width=1.0\textwidth]{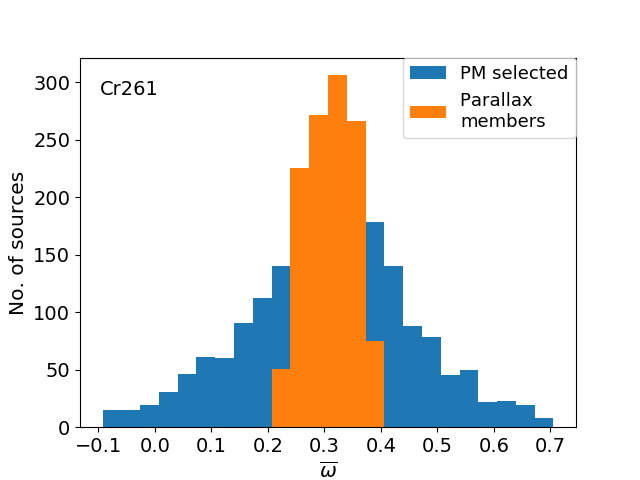}
		\caption*{}
	\end{subfigure}
	\quad 
   	\begin{subfigure}[b]{0.45\textwidth}
   		\includegraphics[width=1.0\textwidth]{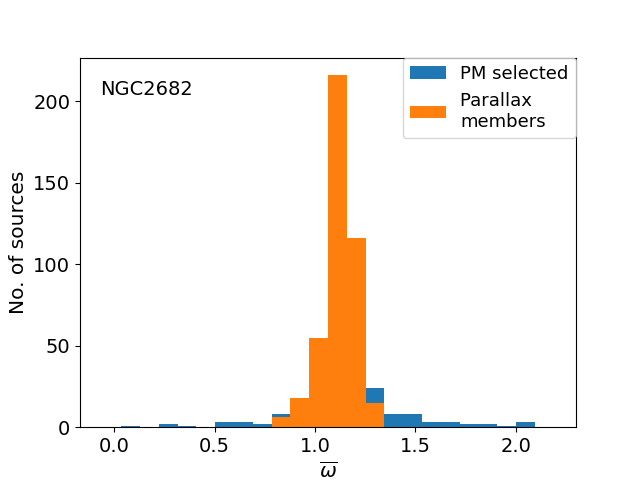}
		\caption*{}
	\end{subfigure}
	\quad
	\begin{subfigure}[b]{0.45\textwidth}
		\includegraphics[width=1.0\textwidth]{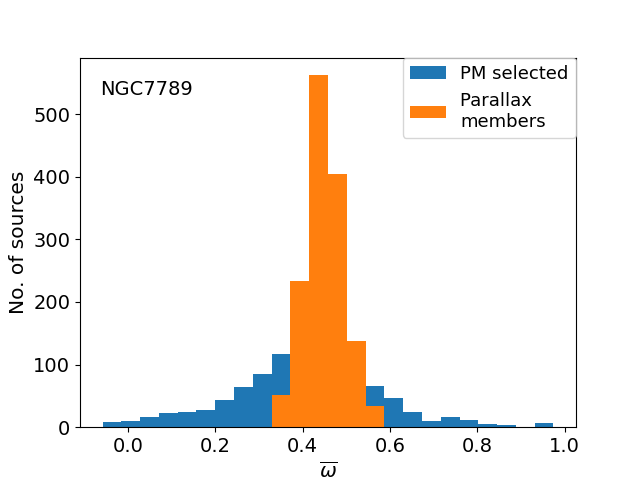}
		\caption*{}
	\end{subfigure}
	\quad
	\begin{subfigure}[b]{0.45\textwidth}
		\includegraphics[width=1.0\textwidth]{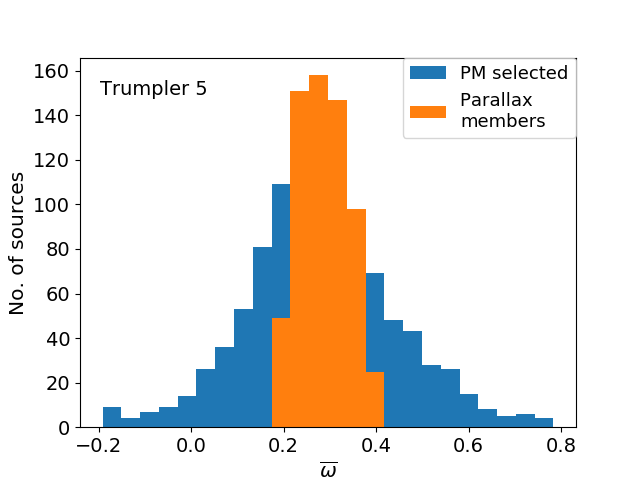}
		\caption*{}
	\end{subfigure}
	\caption{The parallax distribution of the proper motion selected sources (blue histogram) and the proper motion as well as parallax selected sources (orange histogram) of the clusters. The parallax range is selected as $\mathrm{\bar{\omega} \pm 3\Delta\bar{\omega}}$, where $\mathrm{\bar{\omega}}$ and $\mathrm{\Delta\bar{\omega}}$ are the mean of the parallaxes of the proper motion selected sources which are having G <= 15 mag and mean of their parallax errors, respectively, for Collinder 261 and Trumpler 5, and the mean of the parallaxes of the spectroscopically confirmed members and the mean of their parallax errors, respectively, for NGC 2682 and NGC 7789.}
	\label{fig:Figure A2}
\end{figure*}

\begin{figure*}
	\begin{subfigure}[b]{0.45\textwidth}
		\includegraphics[width=1.0\textwidth]{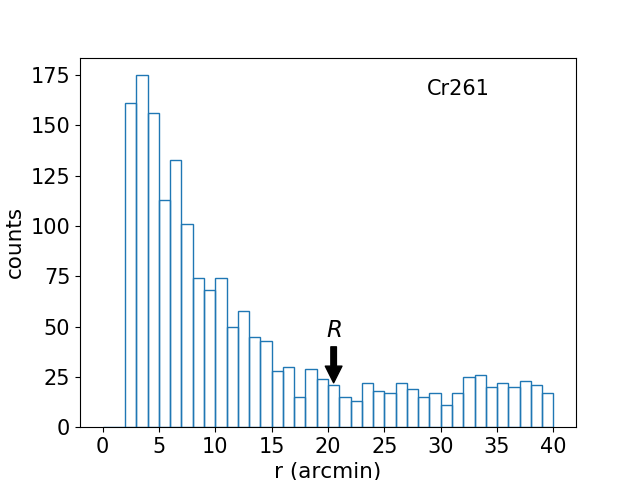}
		\caption*{}
	\end{subfigure}
    	\quad 
    	\begin{subfigure}[b]{0.45\textwidth}
		\includegraphics[width=1.0\textwidth]{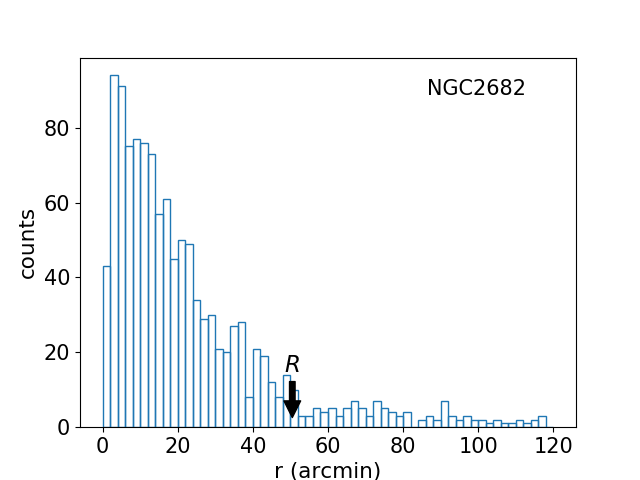}
        	\caption*{}
    	\end{subfigure}
    	\quad
	\begin{subfigure}[b]{0.45\textwidth}
		\includegraphics[width=1.0\textwidth]{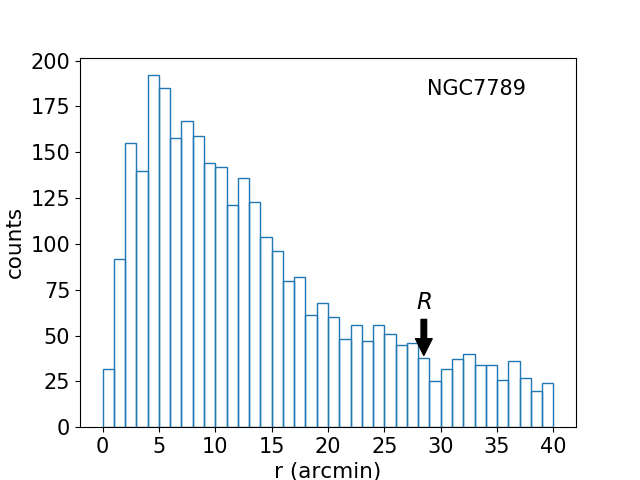}
		\caption*{}
	\end{subfigure}
    	\quad
	\begin{subfigure}[b]{0.45\textwidth}
		\includegraphics[width=1.0\textwidth]{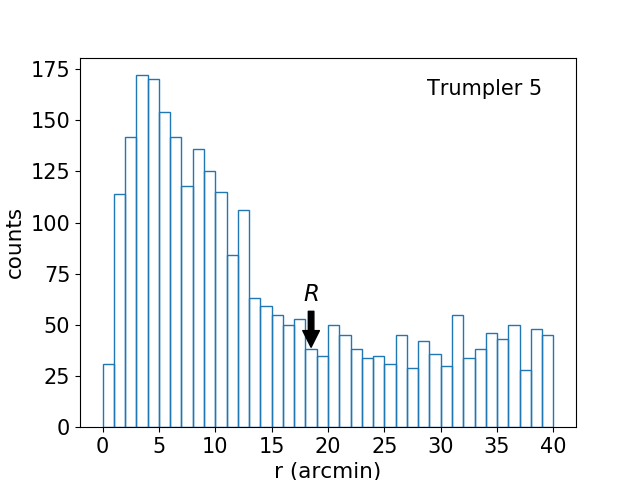}
		\caption*{}
	\end{subfigure}
    \caption{The radial distribution of the proper motion and parallax selected sources. The radius at which the cluster members merge with the field stars is chosen as the cluster radius.}
	\label{fig:Figure A3}
 \end{figure*}

We follow the method developed by \citet{Vaidya2020} to determine the membership of 4 OCs, Collinder 261, NGC 2682, NGC 7789, and Trumpler 5 using the Gaia DR2 data. The method mainly involves three steps: (a) Estimation of proper motion selection range, (b) Estimation of parallax selection range, and (c) Estimation of cluster radius from the sample of sources following the proper motion and parallax selection criteria. For (a), we used two methods, a Gaussian fitting and an application of the mean-shift algorithm on the sources pre-identified as probable cluster members due to their separate proper motions from the field stars, in order to determine the mean values and the spread in the cluster proper motions. Table \ref{tab:Table_A1} lists the mean and the standard deviation of proper motions estimated using the mean-shift method and by fitting the Gaussian function (see Figure \ref{fig:Figure A1}). For (b), we use the parallax values and the corresponding errors of the previously known, spectroscopically confirmed members to fix the parallax selection range for NGC 2682 and NGC 7789, which have many sources studied spectroscopically \citep{Geller2015,Overbeek2014}. For Collinder 261 and Trumpler 5, there are very few sources that have been studied spectroscopically \citep{Mitschang2013,Donati2015}, therefore to fix the parallax ranges in these two clusters, we use those proper motion selected sources which are bright (G <= 15 mag). Figure \ref{fig:Figure A2} shows the over plotted histograms of the parallax and the proper motions selected members and only proper motions selected members. For (c), we plot the radial distribution of sources whose proper motions and parallaxes fall in our selected ranges of the two parameters, for a field of radius that ranges from 40$^\prime$ to 120$^\prime$ around the cluster center. The radius at which cluster members merge with field stars is considered as the cluster radius (see Figure \ref{fig:Figure A3}). The proper motions and parallax selected sources up to the cluster radius are our bonafide cluster members.

\begin{figure*}
	\begin{subfigure}[b]{1.0\textwidth}
		\includegraphics[width=1.0\textwidth]{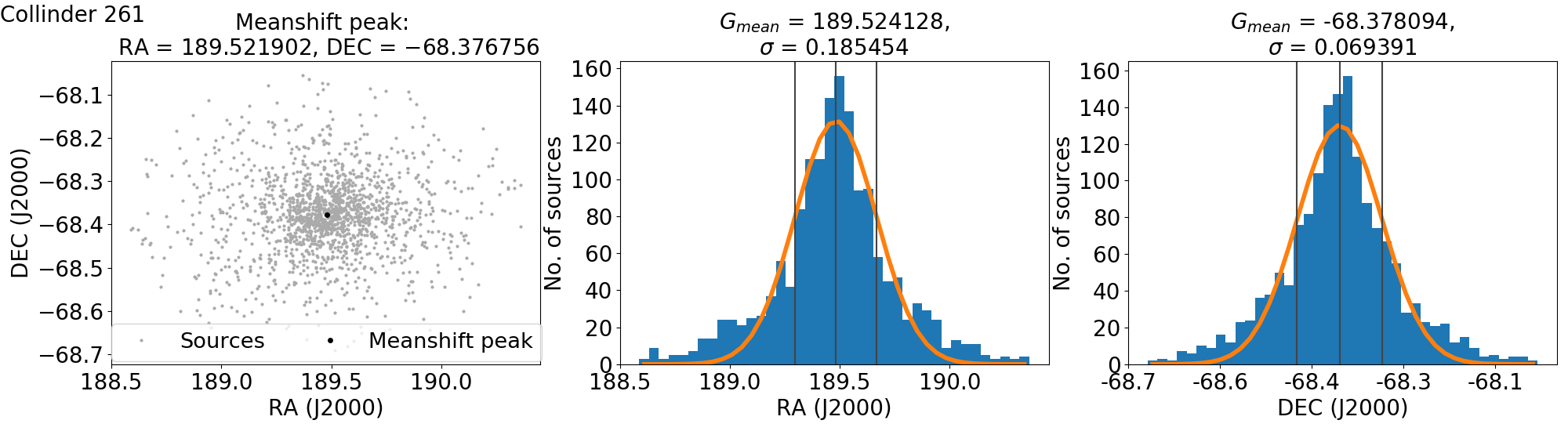}
		\caption*{}
	\end{subfigure}
	\quad
	\begin{subfigure}[b]{1.0\textwidth}
		\includegraphics[width=1.0\textwidth]{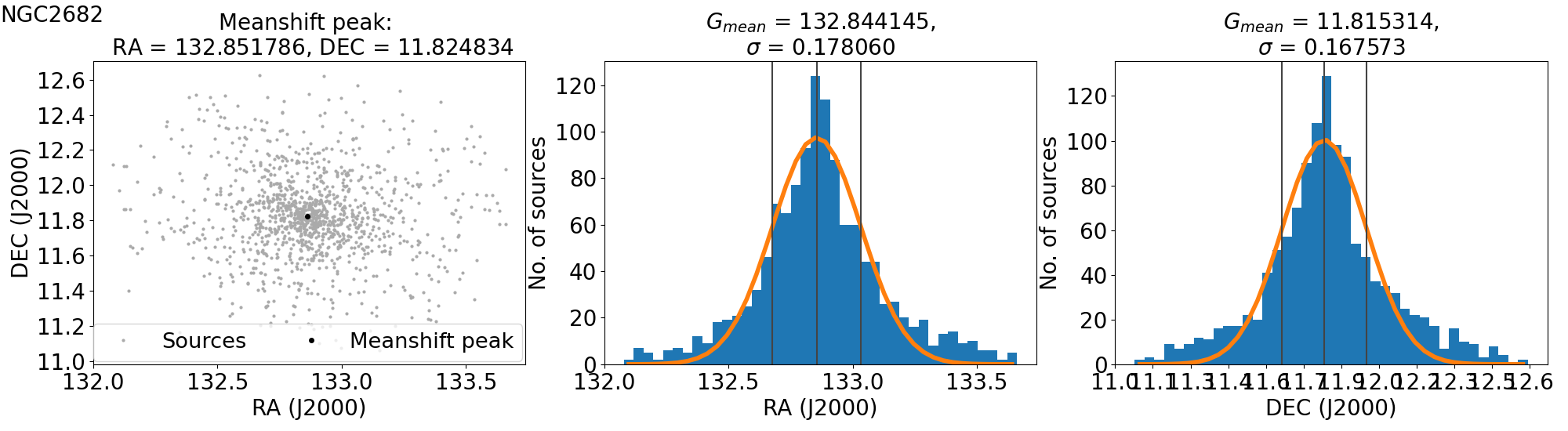}
		\caption*{}
	\end{subfigure}
	\quad
	\begin{subfigure}[b]{1.0\textwidth}
		\includegraphics[width=1.0\textwidth]{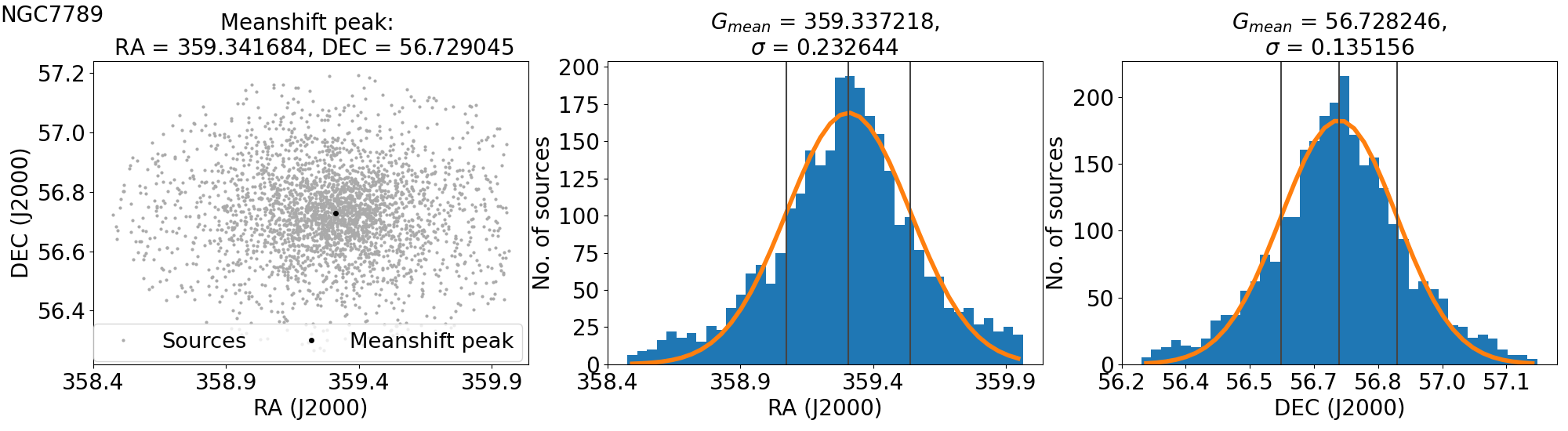}
		\caption*{}
	\end{subfigure}
	\quad
	\begin{subfigure}[b]{1.0\textwidth}
		\includegraphics[width=1.0\textwidth]{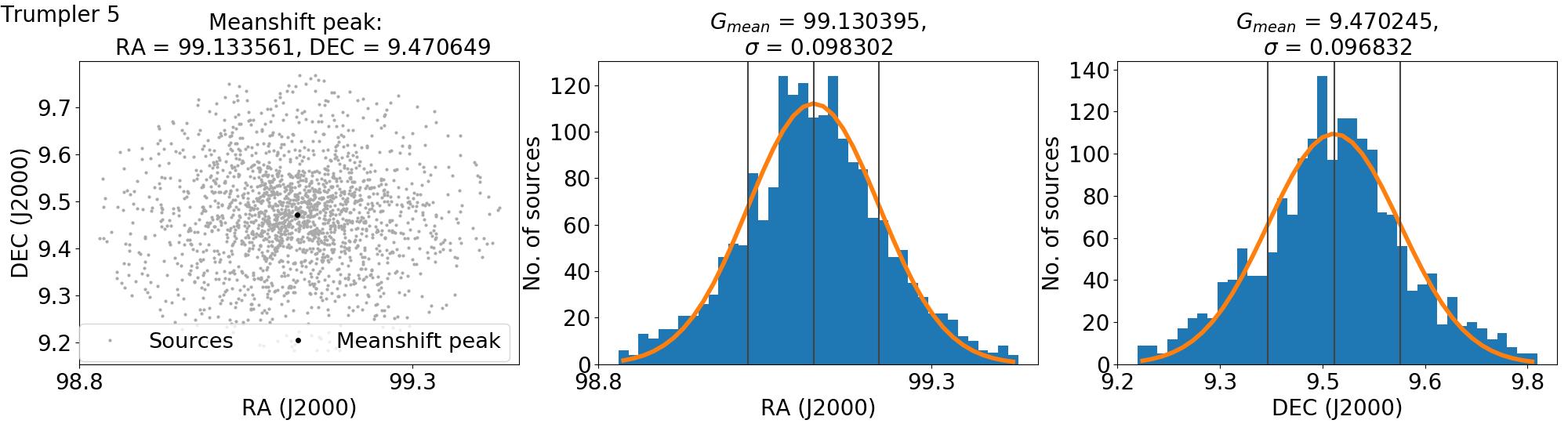}
		\caption*{}
	\end{subfigure}
    \caption{The estimation of cluster centers using the mean-shift algorithm (left panels), and by the fitting of the Gaussian functions to the frequency distributions of RA and DEC (middle and right panels).}
    \label{fig:Figure A4}
\end{figure*}

\begin{figure*}
    \centering
	\begin{subfigure}[b]{0.32\textwidth}
    		\includegraphics[width=1.0\textwidth]{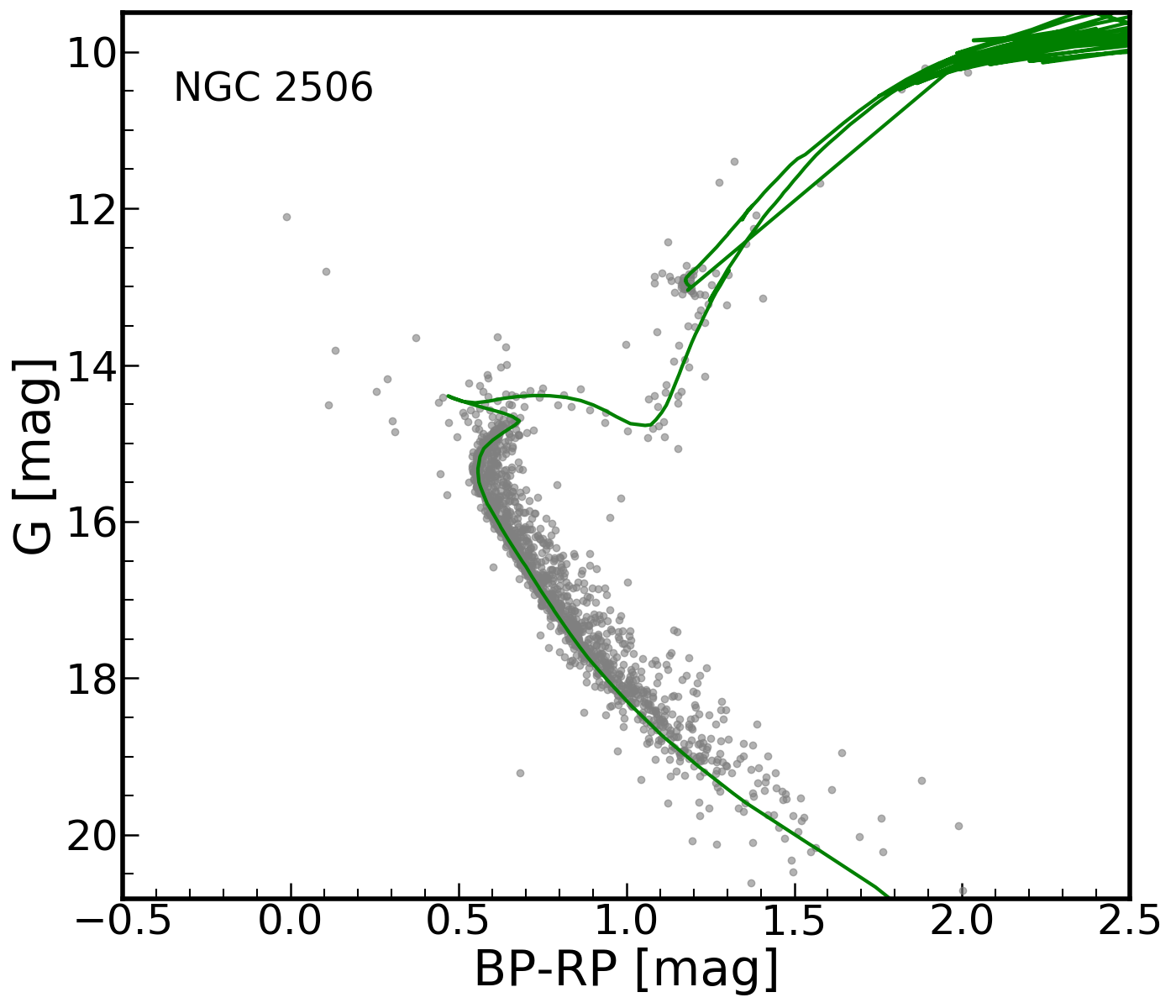}
		\caption*{}
	\end{subfigure}
	\quad 
	\begin{subfigure}[b]{0.32\textwidth}
   		\includegraphics[width=1.0\textwidth]{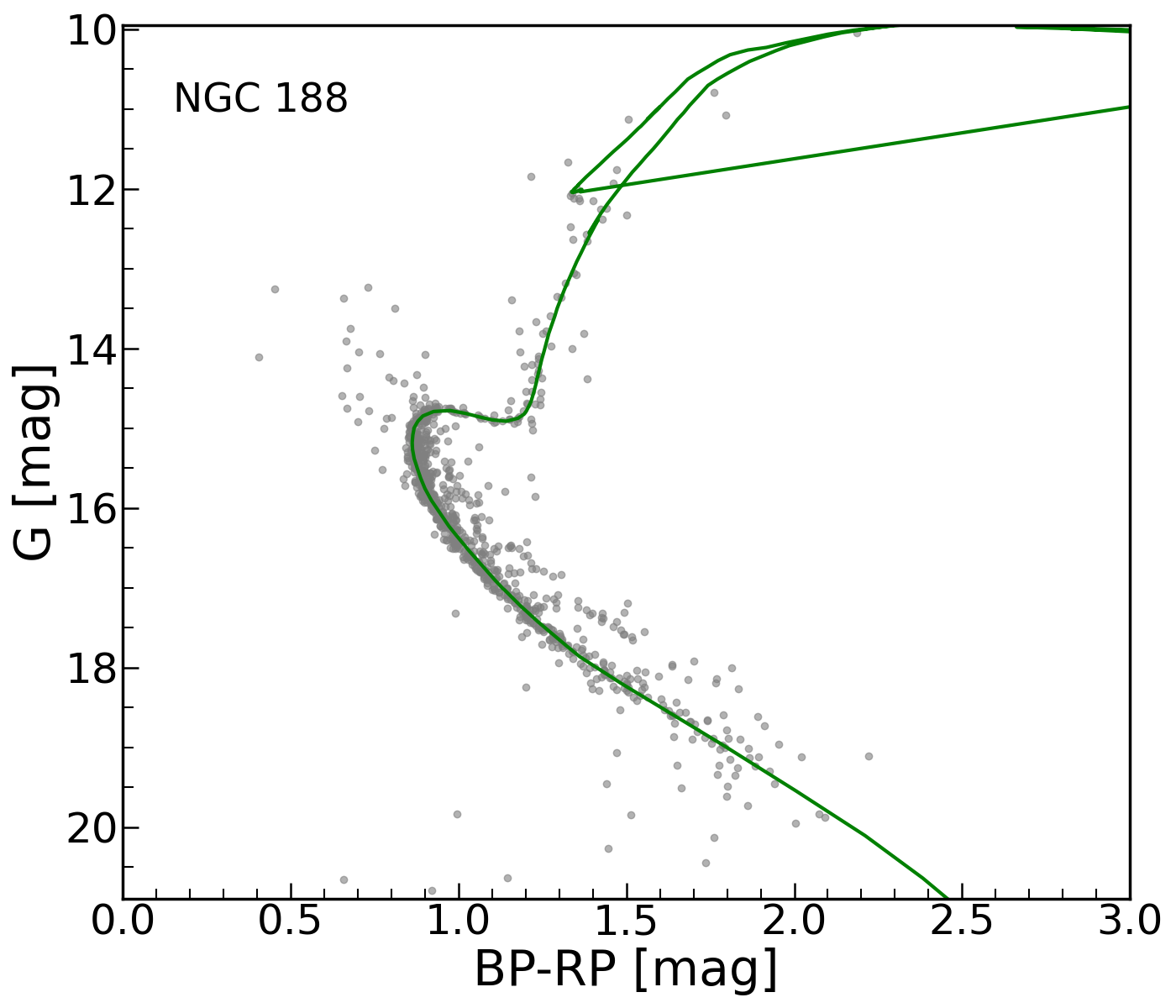}
		\caption*{}
	\end{subfigure}
	\quad
	\begin{subfigure}[b]{0.32\textwidth}
		\includegraphics[width=1.0\textwidth]{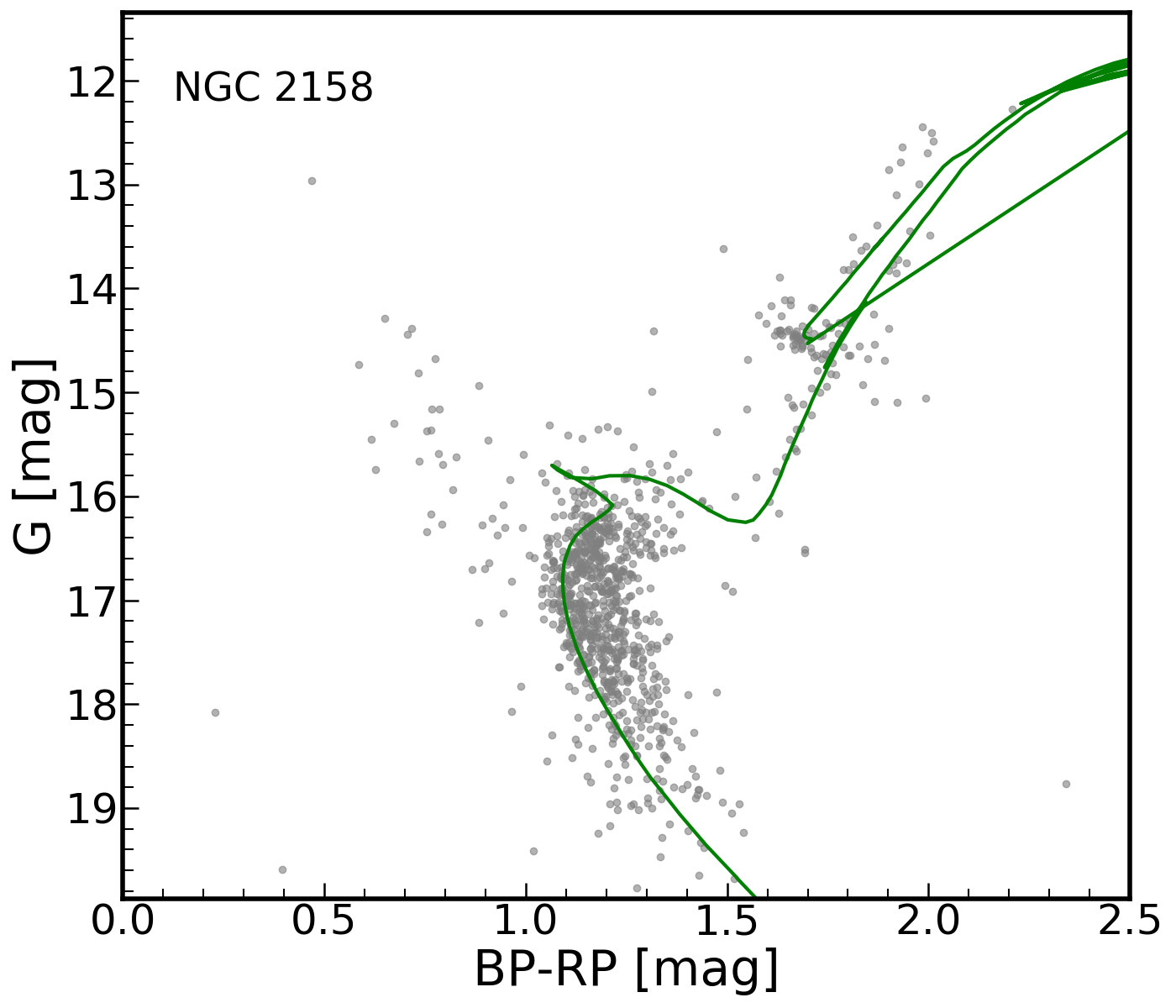}
		\caption*{}
	\end{subfigure}
	\quad
	\begin{subfigure}[b]{0.32\textwidth}
   		\includegraphics[width=1.0\textwidth]{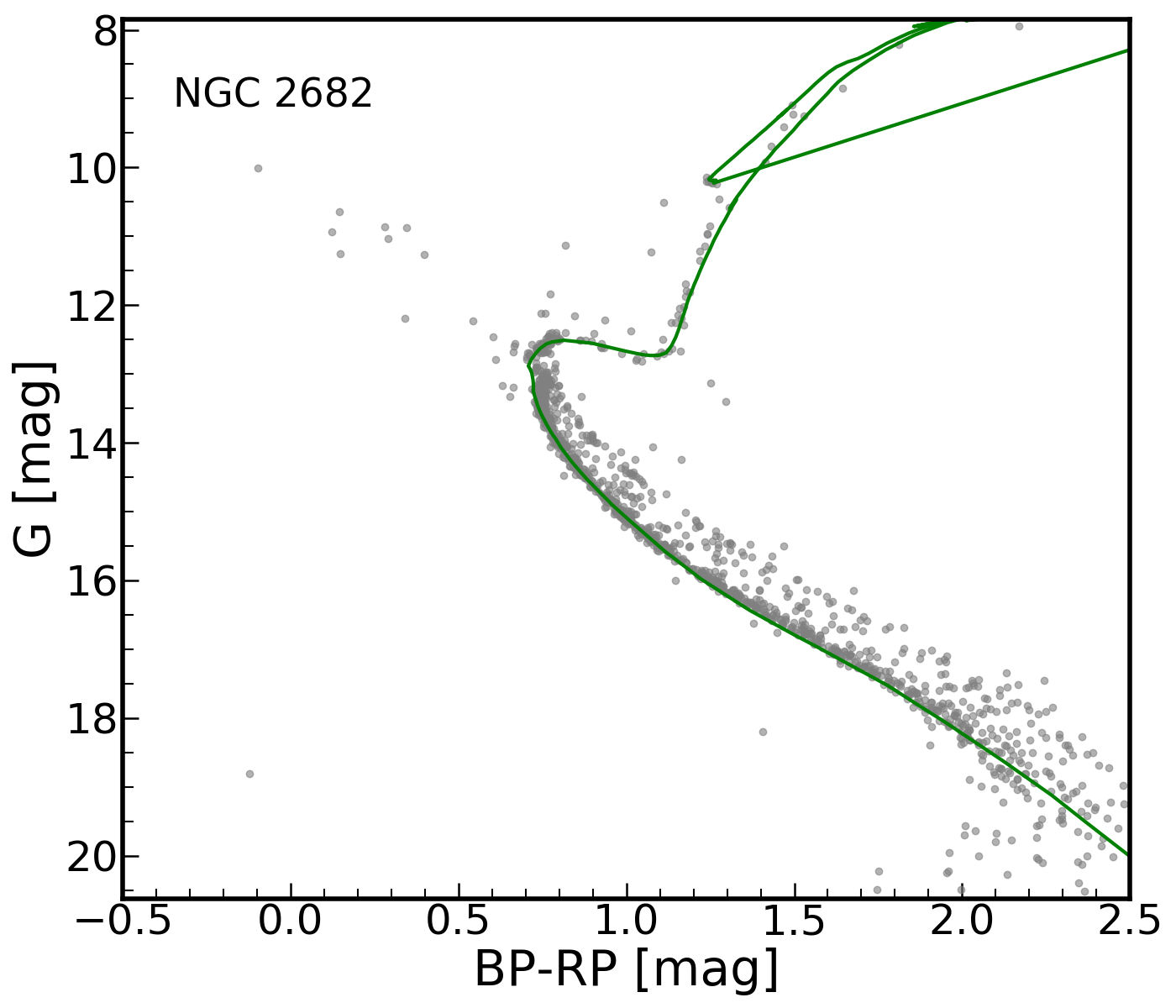}
		\caption*{}
	\end{subfigure}
	\quad
	\begin{subfigure}[b]{0.32\textwidth}
    		\includegraphics[width=1.0\textwidth]{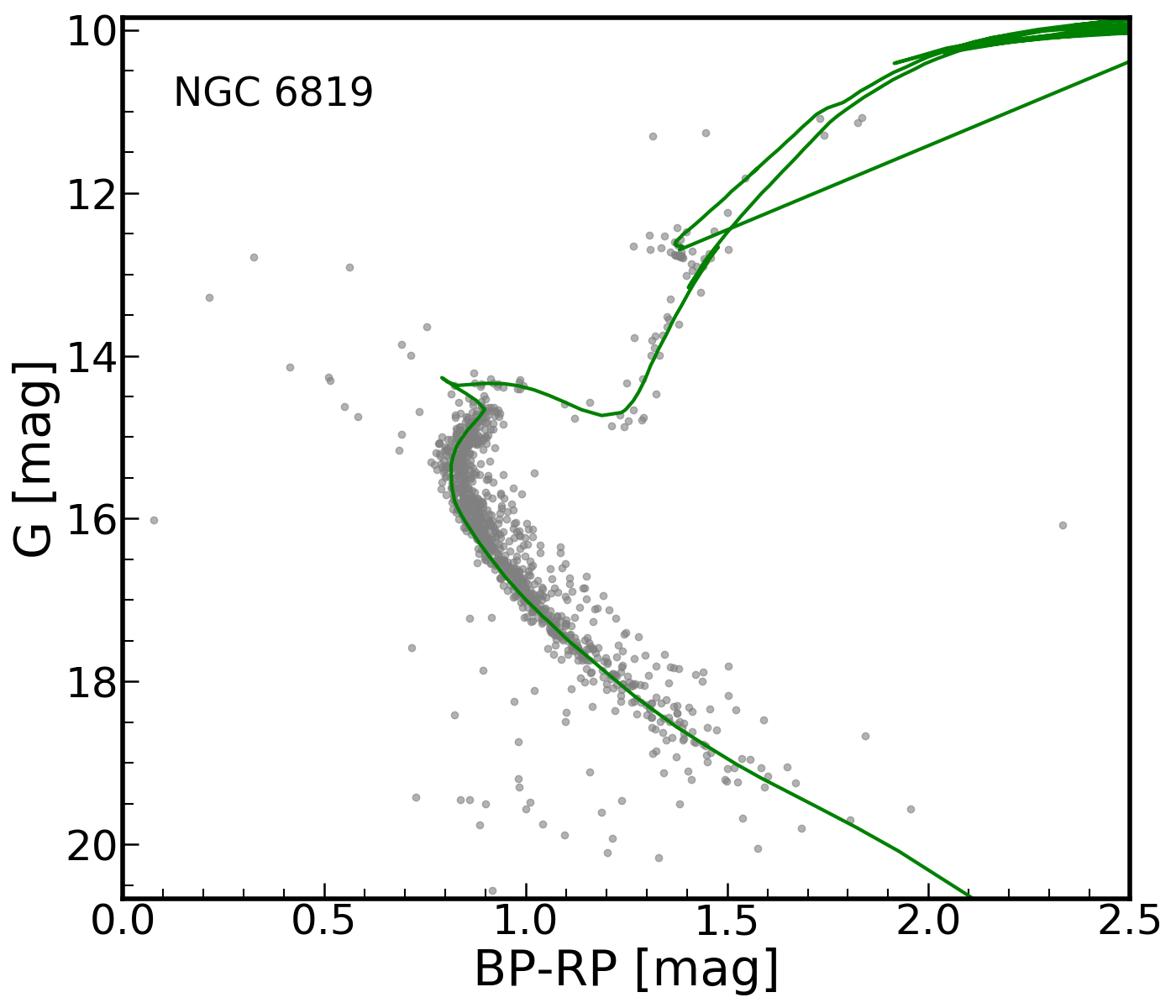}
		\caption*{}
	\end{subfigure}
	\quad 
	\begin{subfigure}[b]{0.32\textwidth}
   		\includegraphics[width=1.0\textwidth]{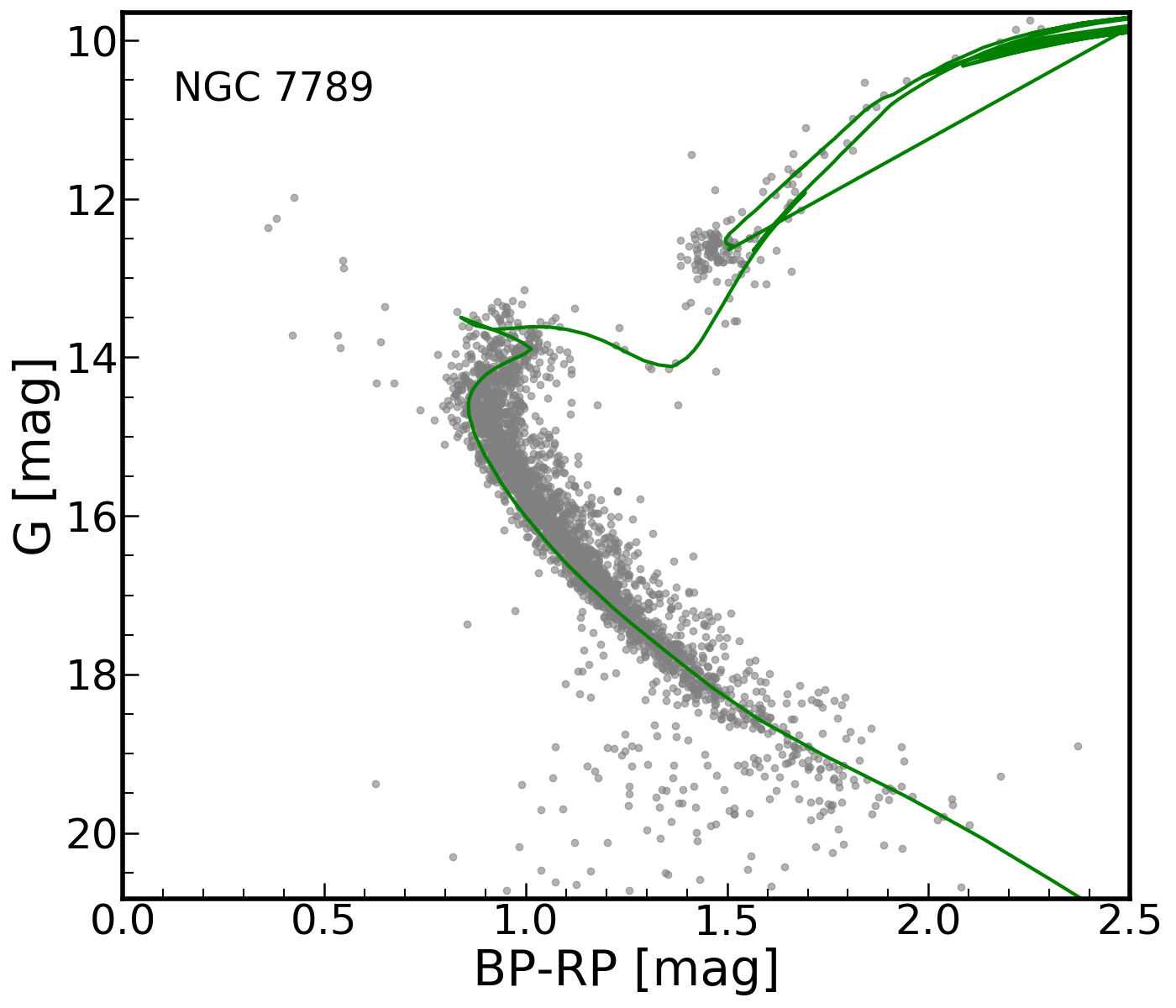}
		\caption*{}
	\end{subfigure}
	\quad
	\begin{subfigure}[b]{0.32\textwidth}
		\includegraphics[width=1.0\textwidth]{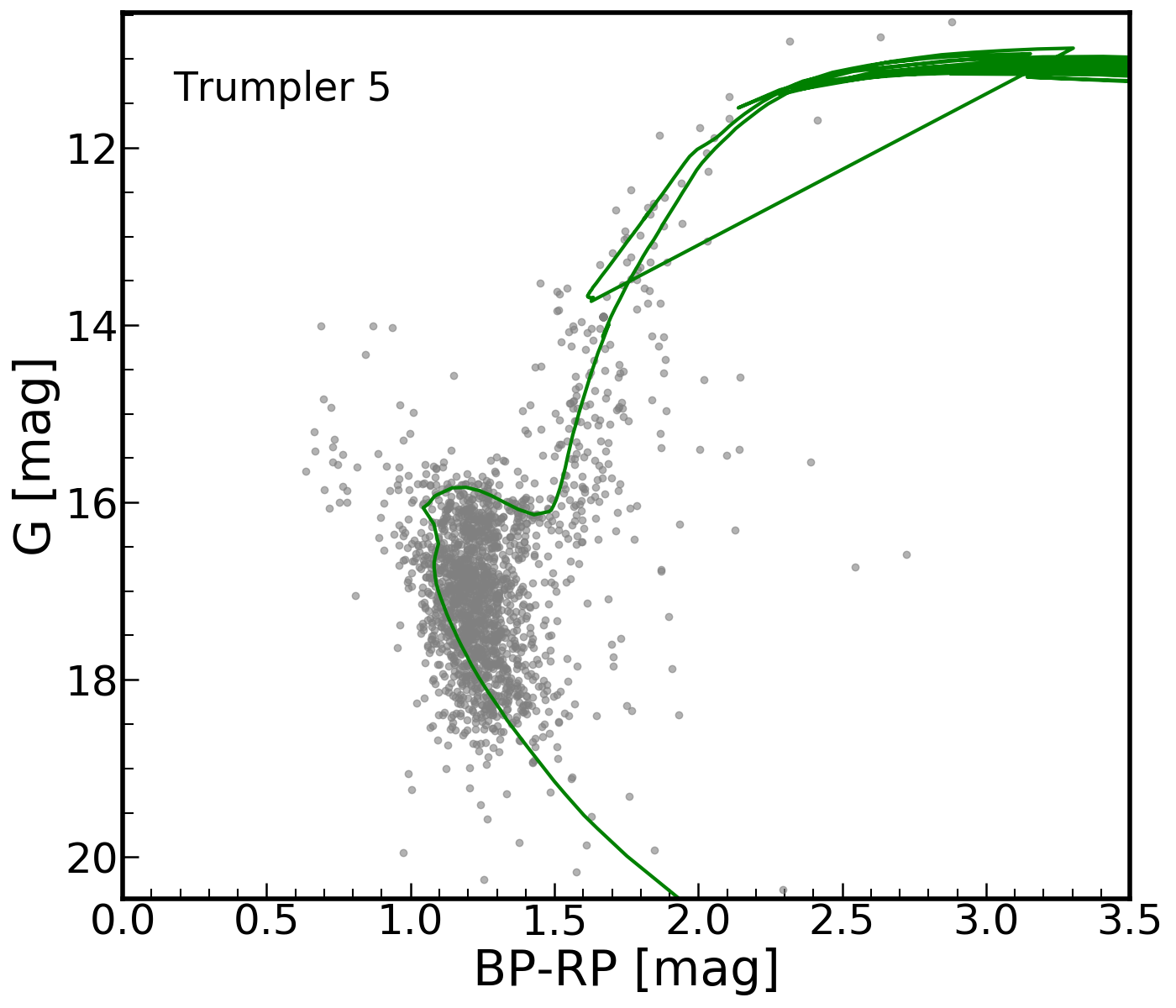}
		\caption*{}
	\end{subfigure}
	\quad
	\begin{subfigure}[b]{0.32\textwidth}
    		\includegraphics[width=1.0\textwidth]{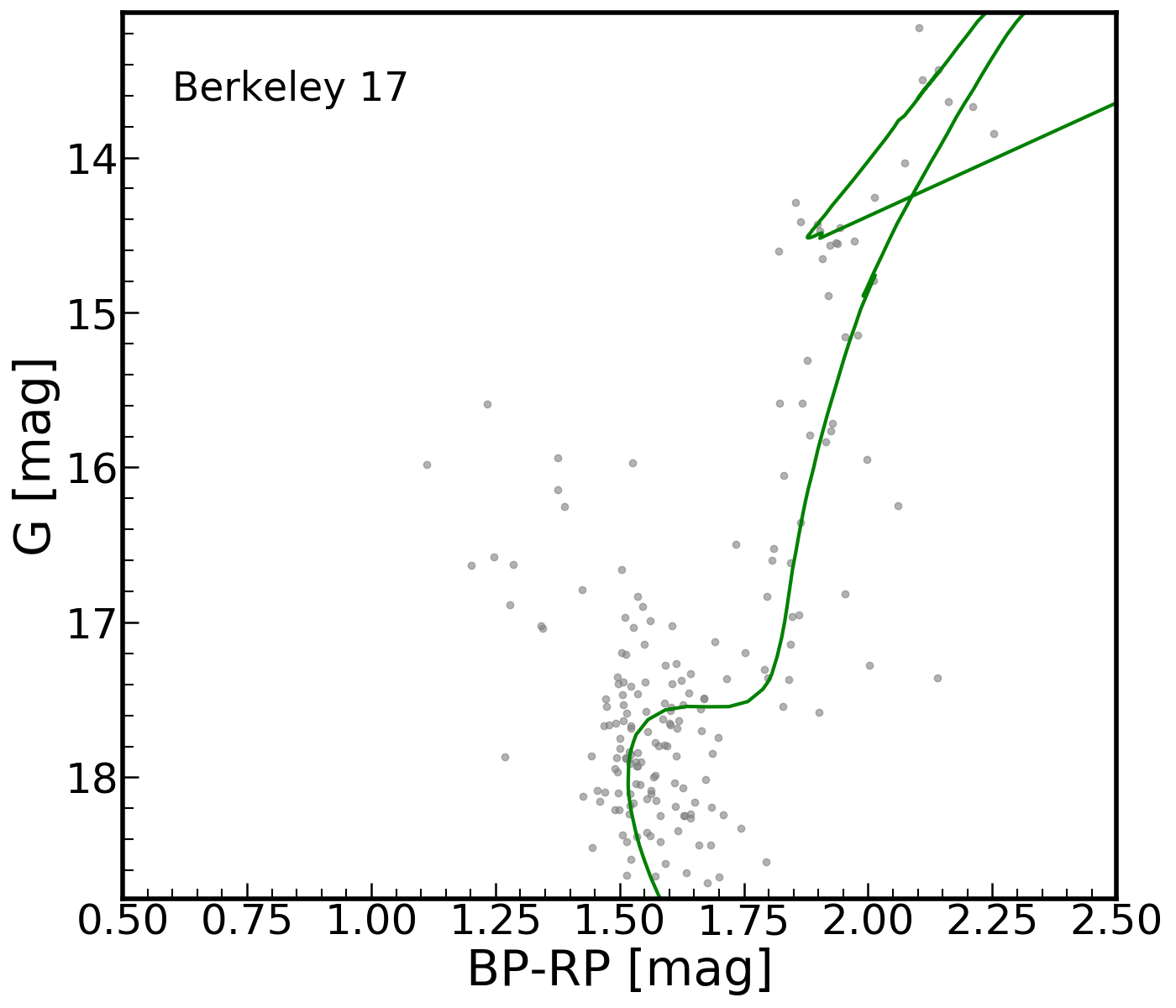}
		\caption*{}
	\end{subfigure}
	\quad 
	\begin{subfigure}[b]{0.32\textwidth}
   		\includegraphics[width=1.0\textwidth]{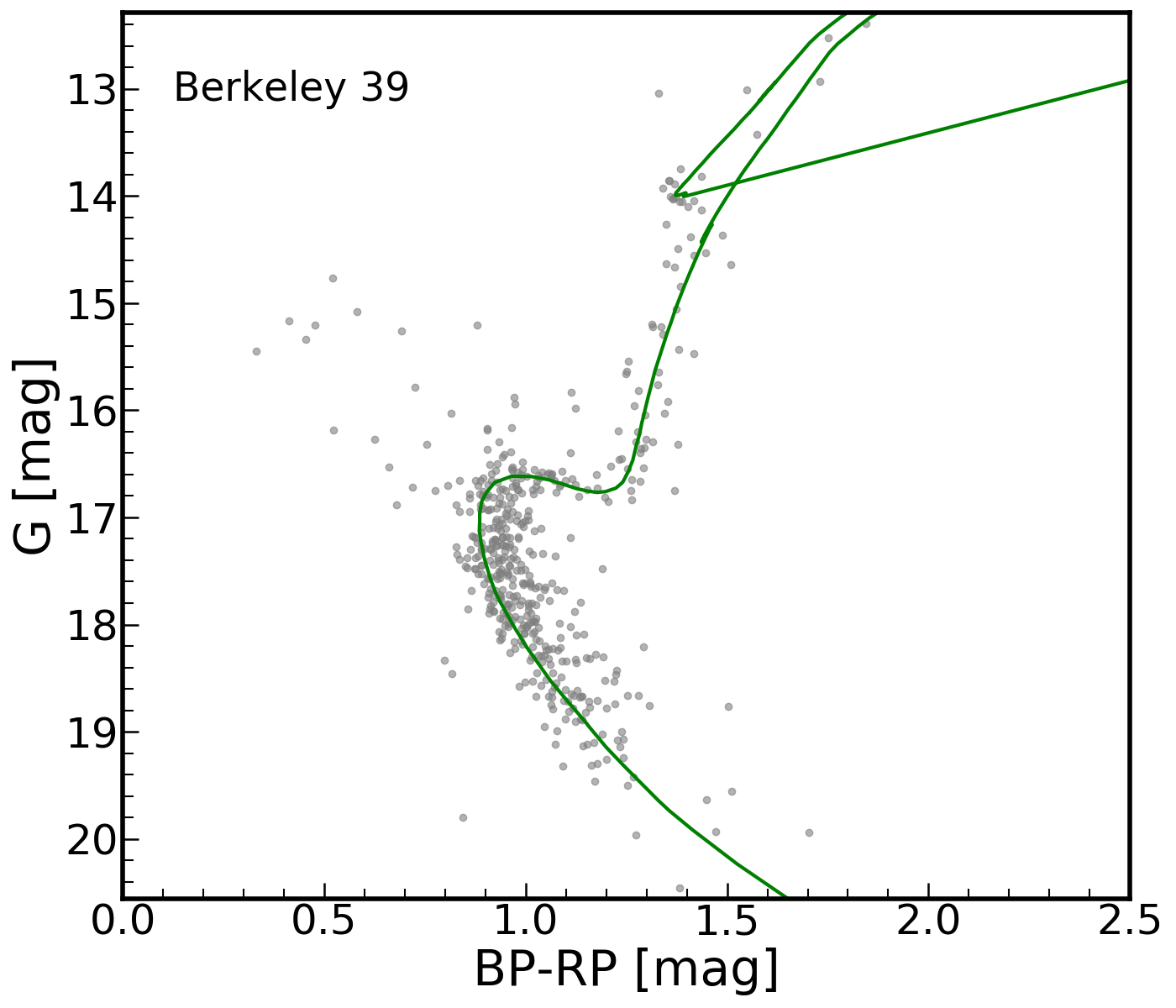}
		\caption*{}
	\end{subfigure}
	\quad
	\begin{subfigure}[b]{0.32\textwidth}
		\includegraphics[width=1.0\textwidth]{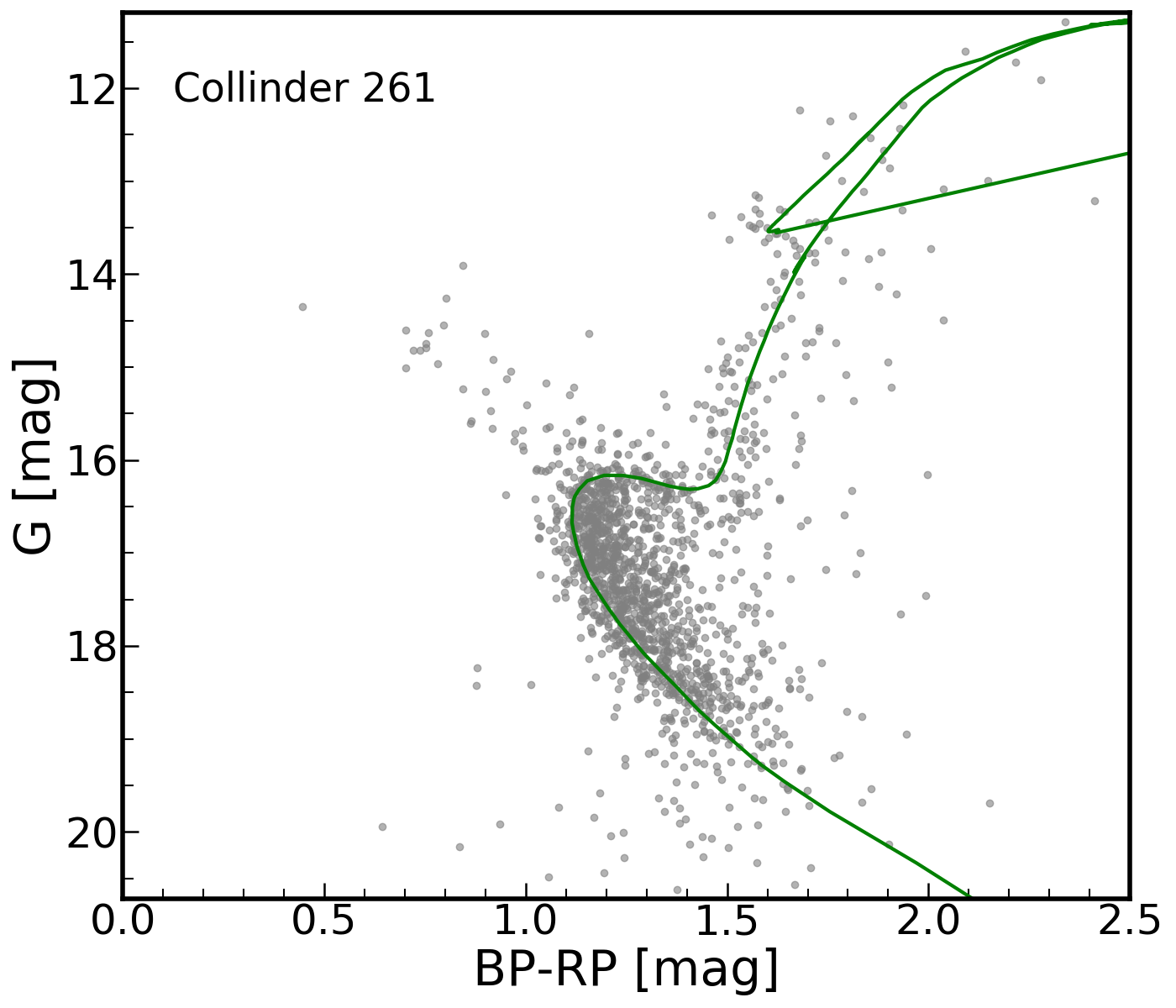}
		\caption*{}
	\end{subfigure}
	\quad
	\begin{subfigure}[b]{0.32\textwidth}
		\includegraphics[width=1.0\textwidth]{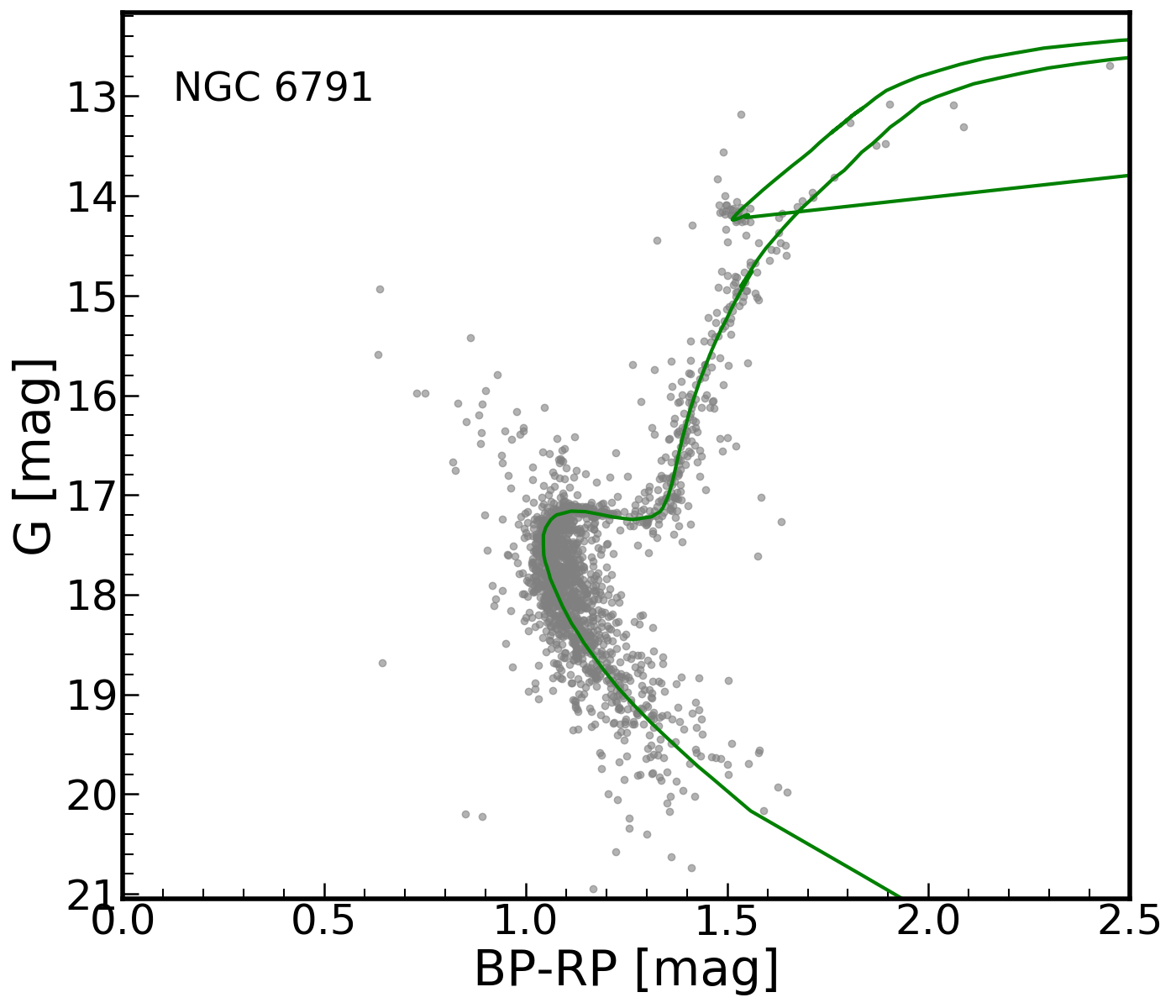}
		\caption*{}
	\end{subfigure}
	\caption{The observed CMDs of all the clusters with fitted PARSEC isochrones as per the parameters listed in Table \ref{tab:Table1}. The CMD shown for Trumpler 5 is obtained after the differential reddening correction}.
	\label{fig:Figure A5}
\end{figure*}

\begin{figure*}
	\begin{subfigure}[b]{0.45\textwidth}
		\includegraphics[width=1.0\textwidth]{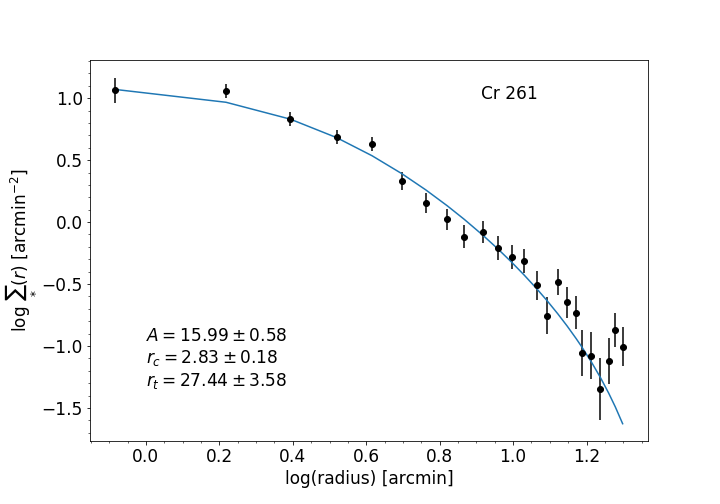}
		\caption*{}
	\end{subfigure}
	\quad 
	\begin{subfigure}[b]{0.45\textwidth}
		\includegraphics[width=1.0\textwidth]{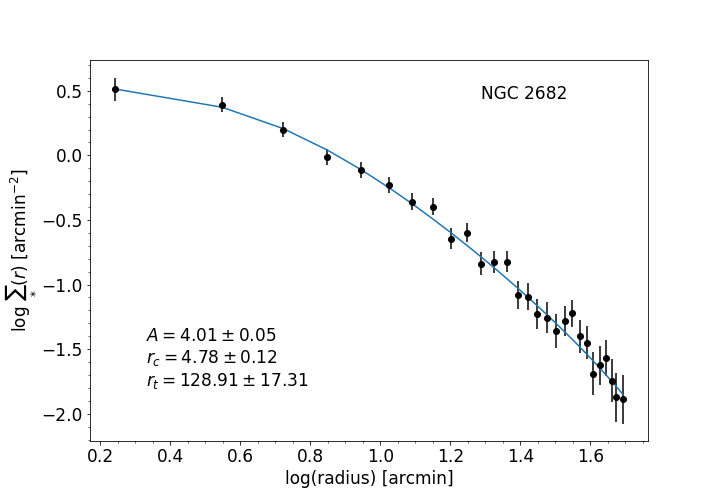}
		\caption*{}
	\end{subfigure}
	\quad 
	\begin{subfigure}[b]{0.45\textwidth}
		\includegraphics[width=1.0\textwidth]{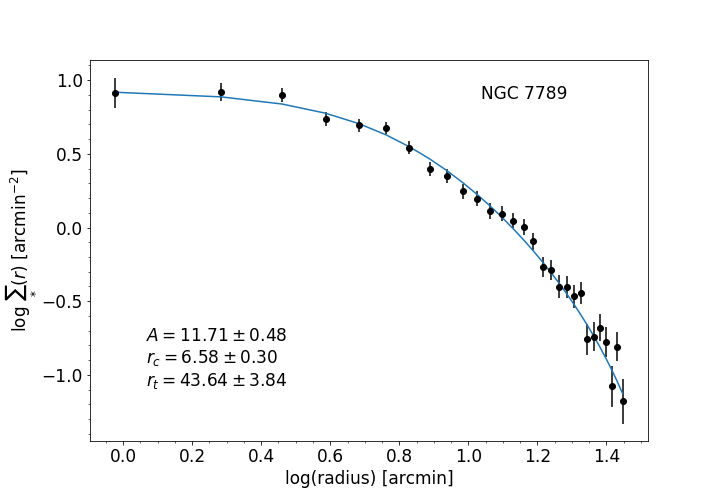}
		\caption*{}
	\end{subfigure}
	\quad
	\begin{subfigure}[b]{0.45\textwidth}
		\includegraphics[width=1.0\textwidth]{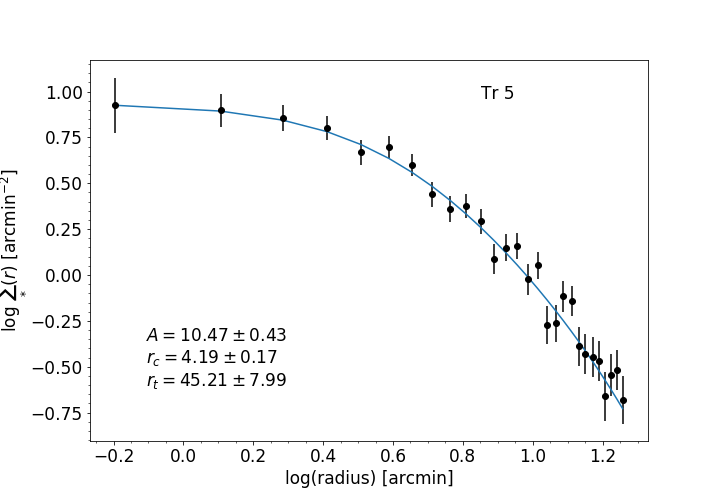}
		\caption*{}
	\end{subfigure}
    \caption{The King's profile fitted to the surface density profile  of the cluster members. The error bars are the $1\sigma$ Poisson errors. The estimated values of the central surface density (A), the core radius ($r_{\mathrm{c}}$), and the tidal radius ($r_{\mathrm{t}}$) of each cluster are marked on the respective plots. A is in the unit of arcmin$^{-2}$ and $r_{\mathrm{c}}$ and $r_{\mathrm{t}}$ are in the units of arcmin.}
    	\label{fig:Figure A6}
\end{figure*}

We then estimate the cluster centers using i) the mean-shift algorithm to find the densest point of the cluster in RA and DEC coordinates \citep{Comaniciu2002} ii) by fitting a Gaussian function to find the mean and the standard deviation of RA and DEC of the cluster members (see Figure \ref{fig:Figure A4}). Table \ref{tab:Table_A3} lists the estimated cluster centers using both of the methods. Figure \ref{fig:Figure A6} shows the fitted King's profiles to the cluster members. The derived values of $r_\mathrm{c}$ and $r_\mathrm{t}$ are listed in column 2 and column 3 of Table \ref{tab:Table2}, respectively. Table \ref{tab:Table_A2} includes the derived fundamental cluster parameters, e.g. age, distance, extinction, color-excess, and metallicity estimated by fitting the Parsec isochrone to the identified cluster members. Figure \ref{fig:Figure A5} shows the observed CMDs and Figure \ref{fig:Figure 1} shows the normalized CMDs of the 11 OCs included in the present work.\\

\pagebreak

\begin{table*}
	\caption{The mean value and the standard deviation of the proper motions in RA and DEC determined from the Gaussian fit, and the peak of the proper motions in RA and DEC estimated by the mean-shift algorithm.}
	\label{tab:Table_A1}
	\begin{tabular}{ccccccc}
     \hline 
     \\
		cluster & $\mathrm{G_{mean}}$ (RA) & $\mathrm{\sigma}$ (RA) & $\mathrm{Peak_{ms}}$ (RA) & $\mathrm{G_{mean}}$ (DEC) & $\mathrm{\sigma}$ (DEC) & $\mathrm{Peak_{ms}}$ (DEC) \\             
		        & mas yr$^{-1}$ & mas yr$^{-1}$ & mas yr$^{-1}$ & mas yr$^{-1}$ & mas yr$^{-1}$ & mas yr$^{-1}$\\
             \\
             \hline
             \\
            Collinder 261 & $-$6.3495 & 0.2207 & $-$6.3496 & $-$2.7078 & 0.2095 & $-$2.7082 \\ 
 
            NGC 2682  & $-$10.9783 & 0.2324 & $-$10.9666 & $-$2.9625 & 0.2417 & $-$2.9623 \\ 
            
            NGC 7789 & $-$0.9301 & 0.1781 & $-$0.9251 & $-$1.9328 & 0.1740 & $-$1.9329 \\ 
 
            Trumpler 5 & $-$0.5621 & 0.2692 & $-$0.5654 & $+$0.2616 & 0.2327 & $+$0.2670 \\  
            \\
			\hline 
	\end{tabular} 
\end{table*}

\pagebreak

\pagebreak

\pagebreak

\begin{table*}
	\caption{ The fundamental cluster parameters from the isochrone fitting of the cluster members and a comparison with the literature values.}
	\label{tab:Table_A2}
	\begin{tabular}{cccccccccc}
		\hline
		\\
		~&~&~&~&This Work&~&~&~&Literature&~\\ \hline
	    Cluster & Age & $\mathrm{d}$  & Radius  & Metallicity &$ A_{\mathrm{G}}$ &$\mathrm{E(Bp-Rp)}$  & Age &$\mathrm{d}$ & Metallicity \\
	    ~&(Gyr)& (parsec) & ($^\prime$) & ($\mathrm{Z}$) & (mag)& (mag)  & (Gyr)&(parsec)&($\mathrm{[Fe/H]}$)\\
		\\
		\hline
		\\
 Collinder 261 & 6.0 & 3053 & 20 & 0.0127 & 0.6 & 0.41 & 6 -- 11 & 2.7 -- 2.9 & $-$0.22 -- $-$0.13  \\
 
 NGC 2682 & 4.0 & 850 & 50 & 0.01 & 0.15 & 0.105 &  3.45 -- 4.8 & 751 -- 891  & $-$0.03 -- $-$ 0.05 \\
 
 NGC 7789 & 1.9 & 1965 & 28 & 0.012 & 0.69 & 0.375 & 1.5 -- 1.8 & 1795 -- 2200 & $-$0.18 -- $+$ 0.04  \\

 Trumpler 5 & 3.4 & 3226 & 18 & 0.01 & 0.78 & 0.48 &  2.4 -- 5.67 & 2800 -- 3080 & $-$0.40 -- $-$0.44 \\
		\\
		\hline
	\end{tabular}
	\begin{tablenotes}
		\item {Collinder 261: Age -- \citet{Carraro1999,Bragaglia2006,Gozzoli1996}, distance -- \citet{Bragaglia2006,Cantat2018,Gao2018}, metallicity -- \citet{Friel1995,Friel2003,Carretta2005,De2007,Sestito2008}}
		\item {NGC 2682: Age -- \citet{Sun2020,Yadav2008,Netopil2016,Bossini2019}, distance -- \citet{Eggen1964,Sarajedini2009}, metallicity -- \citet{Friel1993,Tautvai2000,Jacobson2011,Overbeek2016,Sun2020}}
		\item {NGC 7789: Age -- \citet{Brunker2013,Gim1998,Salaris2004}, distance -- \citet{Gao2018,Wu2009,Gim1998,Cantat2018}, metallicity -- \citet{Friel1993,Chen2011,Jacobson2011,Overbeek2015,Pancino2010}}
		\item {Trumpler 5: Age -- \citet{Kim2003,Salaris2004}, distance -- \citet{Donati2015,Kaluzny1998}, metallicity -- \citet{Donati2015,Netopil2016}}        
	\end{tablenotes}
\end{table*}

\pagebreak


\pagebreak

\begin{table*}
	\caption{The cluster center coordinates determined by the mean shift algorithm and the fitting of the Gaussian function to frequency distributions of RA and DEC of the cluster members.}
	\label{tab:Table_A3}
	\begin{tabular}{ccccc}
     \hline 
     \\
		cluster & RA (deg) & DEC (deg) & RA (deg) & DEC (deg)   \\ 
            
		        & Mean Shift & Mean Shift & Gaussian & Gaussian   \\
             \\
             \hline
             \\
            Collinder 261 & $\mathrm{+189.5219}$ & $\mathrm{-68.3768}$ & $\mathrm{+189.5241 \pm 0.1854}$ & $\mathrm{-68.3781 \pm 0.0694}$ \\ 
 
            NGC 2682 & $\mathrm{+132.8518}$ & $\mathrm{+11.8248}$ & $\mathrm{+132.8441 \pm 0.1781}$ & $\mathrm{+11.8153 \pm 0.1676}$ \\ 
            
            NGC 7789 & $\mathrm{+359.3417}$ & $\mathrm{+56.7290}$ & $\mathrm{+359.3372 \pm 0.2326}$ & $\mathrm{+56.7282 \pm 0.1352}$ \\ 
 
            Trumpler 5 & $\mathrm{+99.1336}$ & $\mathrm{+9.4706}$ & $\mathrm{+99.1304 \pm 0.0983}$ & $\mathrm{+9.4702 \pm 0.0968}$  \\ 
            \\
			\hline 
	\end{tabular} 
\end{table*}

\pagebreak


\pagebreak
\begin{figure*}
	\begin{subfigure}[b]{0.45\textwidth}
		\includegraphics[width=1.0\textwidth]{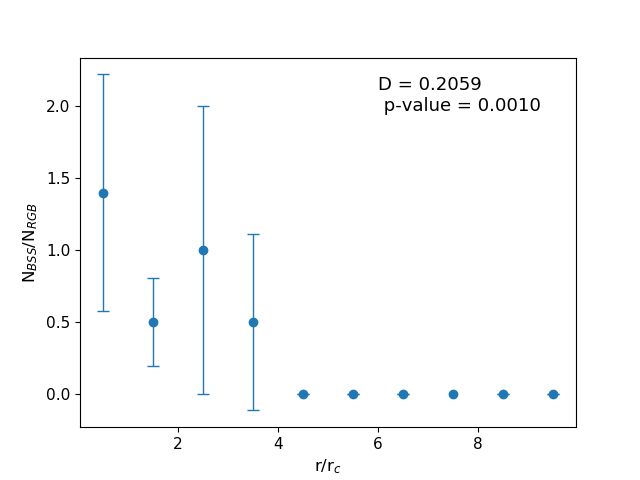}
		\caption*{}
	\end{subfigure}
	\quad 
	\begin{subfigure}[b]{0.45\textwidth}
		\includegraphics[width=1.0\textwidth]{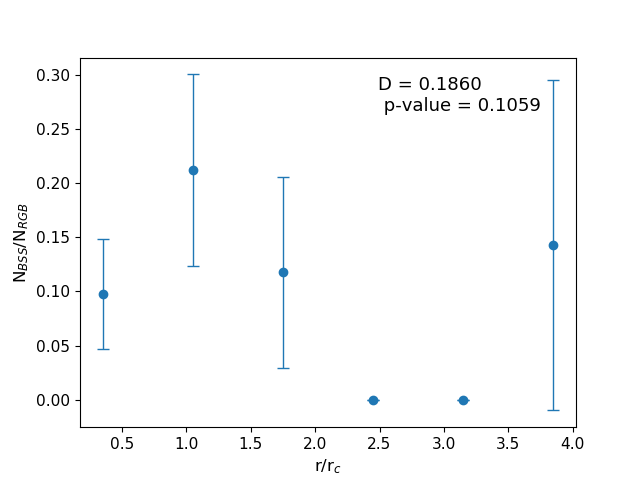}
		\caption*{}
	\end{subfigure}
    \caption{The ratio $N_{\mathrm{BSS}}/N_{\mathrm{RGB}}$ is plotted against the radial distance in the units of $r_{\mathrm{c}}$, for NGC 2682 (left panel) and NGC 7789 (right panel). To plot this, BSS and RGBs of the same magnitude range have been used. The error bars are estimated using propagation of errors. The dip statistic, D, and the p-value estimated from the dip test for bimodality are marked on the plots.} 
	\label{fig:Figure A7}
\end{figure*}
\section{Differential reddening correction in Trumpler 5} 
\label{section:Dr-Tr5}
\begin{figure*}
	\begin{subfigure}[b]{0.3\textwidth}
	\includegraphics[width=1.0\textwidth]{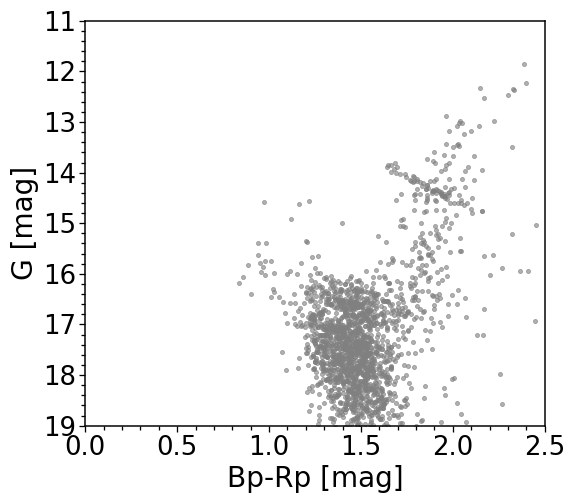}
		\caption*{}
	\end{subfigure}
	\quad
	\begin{subfigure}[b]{0.3\textwidth}
	\includegraphics[width=1.0\textwidth]{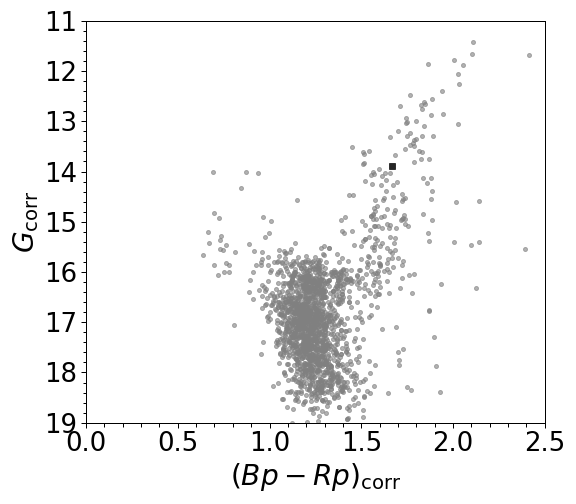}
		\caption*{}
	\end{subfigure}
	\hspace{-1cm} 
	\begin{subfigure}[b]{0.4\textwidth}
	\includegraphics[width=1.0\textwidth]{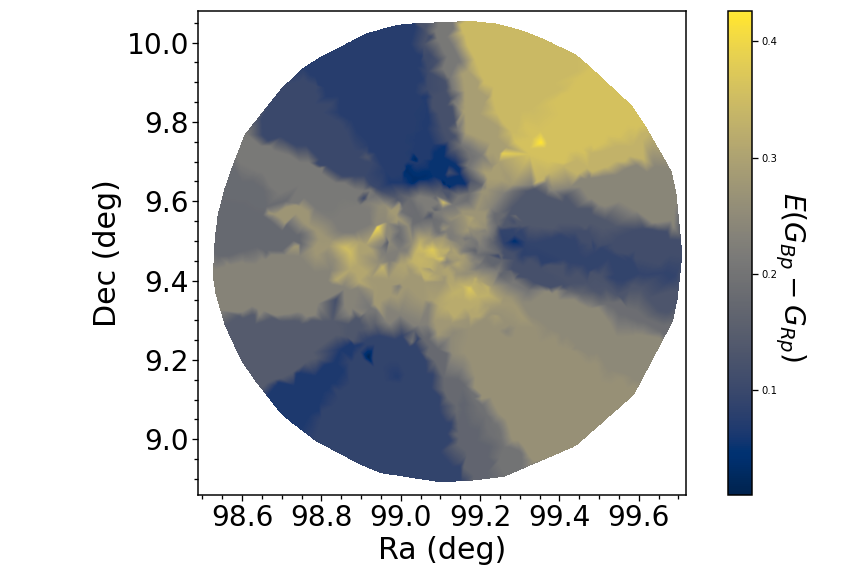}
		\caption*{}
	\end{subfigure}
	\caption{Differential reddening correction in Trumpler 5. The left panel shows the uncorrected CMD of the member stars. The middle panel shows the corrected CMD of the member stars with the black square denoting  the position of the adopted reference point to perform the differential reddening and extinction correction. The right panel shows the differential reddening map of the field of the cluster within $30 \arcmin$ from the cluster center.}
	\label{fig:Figure B1}
\end{figure*}

Among our selected 11 OCs, Trumpler 5 has a broad main-sequence and elongated red clump stars that show that the cluster is highly affected by dust along its line of sight (see left panel of Figure \ref{fig:Figure B1}). Differential reddening correction for this cluster has been recently done by \citet{Rain2020b}. We followed their method to perform differential reddening and extinction correction on our identified cluster members. We briefly describe the method here, for details of the method, readers are referred to section 2.1.1 of \citet{Rain2020b}. First, we selected the red clump stars and calculated the reddening law $R_G = A_G/E(G_{Bp}-G_{Rp})$. We get R$_{\mathrm{G}}$ = 1.73$\pm$0.10 which is equal to R$_{\mathrm{G}}$ = 1.79$\pm$0.05 estimated by \citet{Rain2020b} within the errors. We then choose the same arbitrary point along the reddening line as chosen by \citet{Rain2020b} at G = 13.90 mag and Bp-Rp = 1.67 mag and consider it a zero correction point. We calculated the vertical and horizontal distance of each RC star from this reference point and called it differential extinction, $A_G$ and differential reddening, $E(G_{Bp}-G_{Rp})$, respectively. We then calculated the mean $A_G$ and the mean $E(G_{Bp}-G_{Rp})$ of three nearest RC stars for each cluster member as well as non-members and subtracted it from the magnitudes and colors of the stars. The resulting magnitude and color of the stars are the corrected magnitude and color of the stars. The mean $A_G$ and the mean $E(G_{Bp}-G_{Rp})$ of each star estimated from three nearest RC stars $A_G$ and $E(G_{Bp}-G_{Rp})$ are the differential extinction and differential reddening of the star, respectively. The left panel of Figure \ref{fig:Figure B1} shows the CMD of cluster member before applying the differential reddening and extinction correction, i.e., uncorrected CMD, the middle panel shows the corrected CMD, and the right panel shows the differential reddening map in the field of the cluster. 


\bsp	
\label{lastpage}
\end{document}